\documentclass[fleqn]{article}

\usepackage[T1]{fontenc}
\usepackage[utf8]{inputenc}
\usepackage[english]{babel}
\usepackage{xcolor}
\usepackage{graphicx}
\usepackage{authblk}
\usepackage{textcomp}
\usepackage{titlesec}
\usepackage{hyperref}
\hypersetup{
    colorlinks  =true,
    linkcolor   =blue,
    filecolor   =blue,      
    urlcolor    =blue,
    citecolor   =blue,
   }
\usepackage{cite}
\usepackage[normalem]{ulem}
\usepackage[hmargin=3.5cm]{geometry}

\titlespacing*{\section}{0pt}{0pt}{0pt}
\titlespacing*{\subsection}{0pt}{0pt}{0pt}
\titlespacing*{\subsubsection}{0pt}{1.5pt}{ -.5pt}

\definecolor{richblue}{rgb}{0.19, 0.55, 0.91}

\definecolor{richgreen}{rgb}{0.0, 0.5, 0.0}

\definecolor{round4}{rgb}{.2, 0.2, 0.6}
\definecolor{round3}{rgb}{.2, 0.2, 0.6}
\definecolor{round2}{rgb}{.2, 0.2, 0.6}
\definecolor{round1}{rgb}{0.82, 0.1, 0.26}

\definecolor{VS}{rgb}{0.8, 0.0, 0.0}

\title{Interferometric Scattering (iSCAT) Microscopy \\\& Related Techniques}
\author{Richard W. Taylor}
\author{Vahid Sandoghdar}
\affil{Max Planck Institute for the Science of Light, 91058 Erlangen, Germany.}
\date{}

\begin{document}
\maketitle

\section*{Abstract}
Interferometric scattering (iSCAT) microscopy is a powerful tool for label-free sensitive detection and imaging of nanoparticles to high spatio-temporal resolution. As it was born out of detection principles central to conventional microscopy, we begin by surveying the historical development of the microscope to examine how the exciting possibility for interferometric scattering microscopy with sensitivities sufficient to observe single molecules has become a reality. We discuss the theory of interferometric detection and also issues relevant to achieving a high detection sensitivity and speed. A showcase of numerous applications and avenues of novel research across various disciplines that iSCAT microscopy has opened up is also presented.

\section{Introduction}
Super-resolving the position of nanoscopic objects to a precision better than the wavelength of light is an important and powerful technology in nanoscience and, in particular, in the rapidly growing field of nanobiology. The forebear to modern super-resolution microscopy, where typically one aims to resolve intricate extended cellular sub-structures, is fluorescence microscopy. With the advent of single-molecule fluorescent spectroscopy and microscopy in the early 1990s, it became possible to extend such measurements to fluorescent labels as small as single dye molecules, quantum dots or single fluorescent proteins \cite{Fernandez-Suarez2008}. Fluorescence as a contrast mechanism, however, brings about several restrictions. These include 1) the use of the label itself, which may introduce artifacts to the interpretation, 2) the limited photoemission, caused by photobleaching and photoblinking as well as 3) saturation which curtails the spatio-temporal resolution and duration of a measurement. Fluorescence-free alternatives are thus highly desirable to overcome these limitations.  

\begin{figure}[!htb]
    \centering
   \includegraphics[scale=.65]{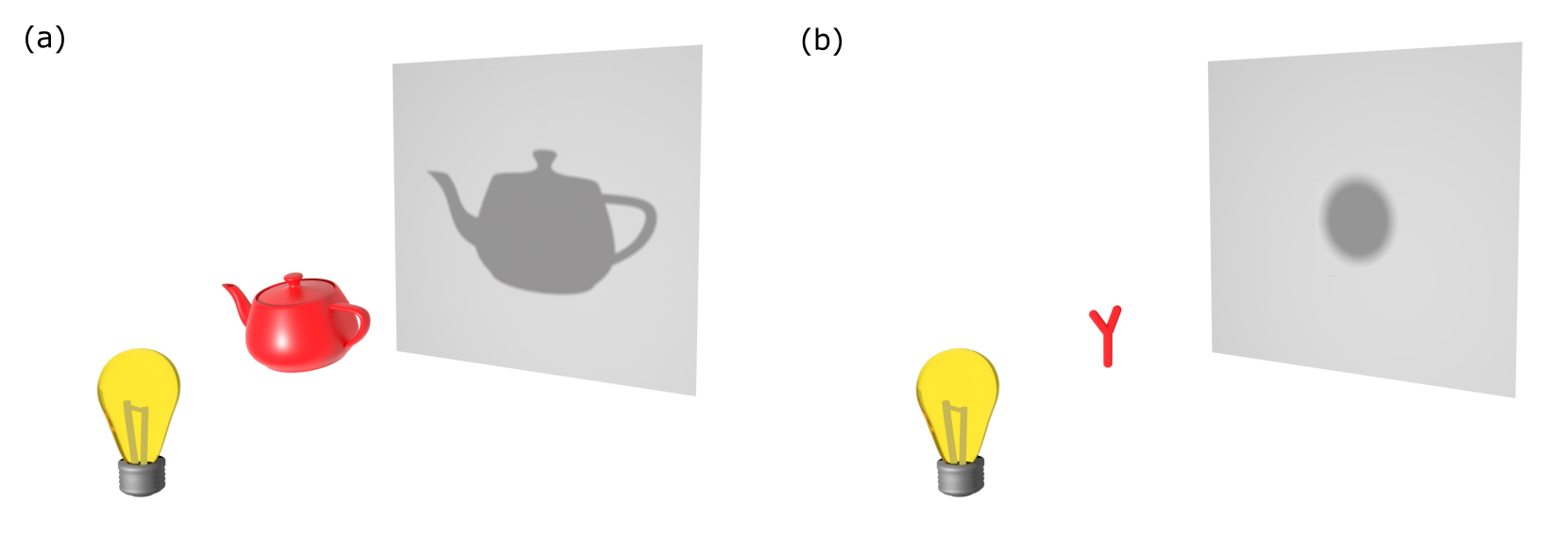}
    \caption{ Detection of an object from its shadow: just as we may observe an everyday object from the shadow its casts - such as the teapot, we may equally do so with an object much smaller than the wavelength of light i.e. a nanoparticle or even a single molecule. Such a small object casts a shadow with no structure, instead the shadow appears as a single spot with a minimum size given by the diffraction limit of light.}
    \label{fig:intro_microscopy}
\end{figure}

The most familiar and common mode of fluorescence-free microscopy is when one detects the optical \textit{shadow} cast by the object (see Figure\,\ref{fig:intro_microscopy}a). The size of the shadow is proportional to the size of the object, and its degree of darkness is a measure for its transparency. This detection principle serves as the foundation for the earliest developments in microscopy and continues to be the basis of any modern microscope. It turns out that this line of thought is also applicable for viewing sub-wavelength nanoscopic particles, and even single molecules. In this regime, the geometric shape and size of the object is no longer represented by the shadow, which collapses to the point-spread function (PSF) of the microscope. Nevertheless, the object imprints its signature in the faint \textit{extinction} of the illuminating beam (see Figure\,\ref{fig:intro_microscopy}b). The challenge lies in reaching a high sensitivity in detection to observe the resulting \textit{nano-shadow}, requiring discrimination of the minuscule changes in light intensity as well as oftentimes measures to separate the desired shadow from the accompanying background. 

\begin{figure}[!htb]
    \centering
   \includegraphics[scale=.6]{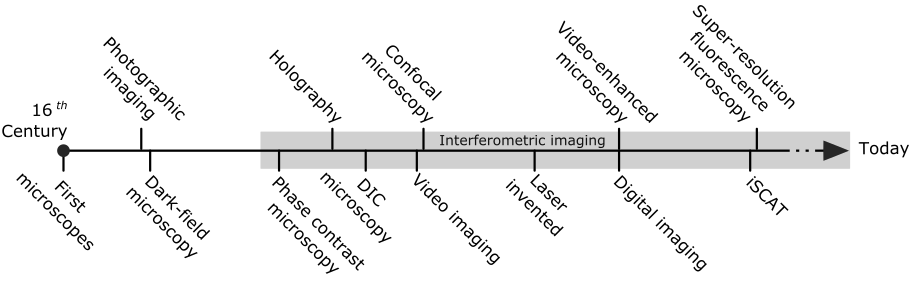}
    \caption{Important milestones throughout the history of microscopy.}
    \label{fig:timeline}
\end{figure}

In this review, we shall show that successful detection of the shadow from nano-objects permits label-free localization with outstanding spatial and temporal resolution, made possible by a high signal-to-noise ratio (SNR). As we shall see, the resulting extinction nanoscopy can be seen as a very recent realization of an old contrast mechanism. It is thus instructive to take a retrospective look at some of the developments in light microscopy that have led us to this exciting possibility, which we summarize in Fig.\,\ref{fig:timeline}.  

\section{Historical perspective}
One of the earliest documented examples of a microscope is attributed to Zacharias Jansen (1585-1638) who reportedly demonstrated a compound two-lens microscope as early as 1590. The microscope boasted a 20-fold magnification and was able to image sufficiently opaque samples when viewed with the eye. About half a century later, both Robert Hooke (1635-1703) and Antonie van Leeuwenhoek (1632-1723) would use microscopes as the foundation for their study into small organisms. Remarkably, the self-made microscopes of the latter reached a magnification one order of magnitude larger than his contemporaries - placing it on the level with current research-grade microscopes, permitting him to observe individual blood and yeast cells, as well as microscopic bacteria and protozoans. These laid the foundations for the discipline of biological microscopy.

In the time following van Leeuwenhoek up to the turn of the twentieth century, the art of microscopy would mature into a rigorous scientific discipline following continuous refinement and innovation. For example, after 1820 it became technically possible to manufacture combinations of lenses small enough for high magnification objectives that could additionally include corrections for chromatic aberrations. In 1873 Ernst Abbe (1840-1905) would publish on his theory of the microscope \cite{Abbe1873}, introducing strict mathematical definitions to the vocabulary of microscopy, which in turn led to better designed instruments. Thereafter, commercial availability of microscopes with good performance led to their widespread adoption by scientists of that era. While during this era we saw a burgeoning diversity in the specimens which became the subject of microscopic investigation, the smallest specimens investigated remained limited to that of micron-sized entities, something already observable since the seventeenth century. To appreciate why such a limit was encountered, it is important to consider the sensitivity of the detector commonly used in these microscopies, namely, the eye. 

The response of the human eye to light intensity is nonlinear, beginning with a detection threshold of around 2\%. Our ability to sense changes in light intensity - which we call \textit{contrast}, diminishes as the change becomes a smaller fraction of the overall brightness \cite{Steinhardt1936}. In conventional bright-field microscopy this means that if small, thin or otherwise weakly-interacting samples cannot cast a shadow of sufficient darkness then, to our eyes, it cannot be detected. An apt example of our contrast sensitivity is that of the spider web \cite{Martin1931}; viewed against the bright sky the web is practically invisible, but seen illuminated by the same sunlight against the darker surface of the ground, the web is now readily observed. In other words, we can circumvent the limitations of our visual sensitivity by diminishing the brightness of the background. This situation is also familiar to the astronomer, as favourable conditions for observing the faint light of distant stars occurs when the surrounding sky is as dark as possible. 

In microscopy, the deliberate removal of background illumination is known as dark-field, with the first recorded instance of its implementation dating back as early as 1830 to Joseph Jackson Lister (1786-1869) \cite{Lister1830}. Despite its effectiveness and advocates, dark-field microscopy initially failed to gain appreciation in the wider circles of microscopy, and was seen somewhat of a novelty. The technique remained largely overlooked \cite{Gage1920} until the realization of the potential for dark-field to observe minute specimens that were otherwise \textit{invisible} in bright-field. Austrian Chemist Richard Adolf Zsigmondy (1865-1929) would beautifully demonstrate this principle in 1902. Zsigmondy, along with Henry Siedentopf (1872-1940), presented the technique for the observation of nanometer-sized gold particles \textit{by eye}. Consisting of an orthogonal objective and condenser, the `ultramicroscope' - as it was so named, allowed the observation of light scattered by the sample in the absence of a background \cite{Siedentopf1902}. This technique facilitated a more sensitive detection microscopy, which would enable, for example, the breakthroughs of Fritz Schaudinn (1871-1906) into the micro-organisms responsible for syphilis. In recognition of his achievements, Zsigmondy would be awarded the Chemistry Nobel Prize in 1925, the first of many Nobel Prizes to be awarded for advances in optical microscopy. It should be noted, however, that dark-field imaging becomes increasingly challenging for smaller objects, especially those in the presence of larger ones, as noise and incomplete removal of the background obscures detectability.

Where dark-field imaging seeks to better detection sensitivity through complete removal of the background, a more powerful approach to better imaging contrast is to selectively shift the phase of the background illumination with respect to the incident light that interacts with the sample. This realization would be made in the decade following Zsigmondy's Nobel Prize by Frits Zernike (1888-1966), who developed such wisdom from the study of aberrations and coherence in diffraction gratings and telescopic optics \cite{Zernike1934,Zernike1938,Zernike1955}. Phase Contrast Microscopy (PCM) \cite{Zernike1942-pt1,Zernike1942-pt2,Zernike1955}, as it became to be known, involved the addition of a phase plate to the illumination path in the microscope with the effect of enhancing the bright-field contrast of specimens with refractive index similar to that of the surrounding media. Phase Contrast Microscopy (PCM) represented a monumental advancement in microscopy, garnering Zernike the 1953 Physics Nobel Prize, with the successes brought about within cancer research being cited as one of the many advances enabled by PCM. Arguably, Zernike's greater impact was to bring interferometric principles to the fore of contemporary microscopy, specifically into biological imaging, where the early success of staining techniques had quashed desires to search for more elaborate means to improve contrast. Another interferometric modality that has since become popular is Differential Interference Contrast (DIC) microscopy, in which regions of differing optical paths in the sample, such as those occurring at edges, are shown with enhanced contrast through the interference of parallel sheared beams. The spatial displacement and angular deviation of the beams, on the order of half the airy disk diameter, is achieved through a special prism introduced by Georges Normarksi (1919-1997)  which bears his name \cite{Normarksi1950}. Interestingly, the conceptual roots of DIC microscopy can be traced to the interferometer developed by Jules Jamin (1818-1886) in 1856 \cite{Jamin1856}, and the first interference microscope from Jacobus Laurens Sirks (1837-1905) in 1893 \cite{Sirks1893}. Another important related invention was that of holography as proposed by Dennis Gabor (1900-1979) in 1947 \cite{Kock1971}, an accomplishment that would lead to Gabor being awarded the Physics Nobel Prize in 1971. Although this new technique was mainly employed in macroscopic imaging, one of the first holographic microscopes was reported in 1966 \cite{Ellis1966}, enabled by the commercial availability of the laser which was invented several years prior. 

By the era of the late 1950s, interferometric microscopy was in its heyday, with a bewildering array of interferometric microscopes reported, several of which found commercial release. A comprehensive review of these microscopes is provided in Ref.\,\cite{Krug-book}. Popular categories of design for these interferometric microscopes include \cite{Shribak-book, Cherry-book}: (1) the beam-shearing type, (2) a two-arm design, and (3) the dual focus type in which the reference wave is focused to a different level than the specimen plane. Unfortunately, interferometric microscopies witnessed a general decline post the 1950s, save for their application in surface profilometry, e.g., in precision engineering, and later for inspection in the microelectronics industry, where reflection-based Normarksi DIC and the Mirau interferometer objective remain workhorses to this day \cite{deGroot2015}. The fall of interferometry as the principle imaging technique in the life sciences was as a result of a series of innovations and developments in fluorescence microscopy \cite{Renz2013, Masters2010}, such as the invention of immuno-labeling in 1950, which led to a meteoric rise in fluorescence imaging, aided by the subsequent invention of confocal scanning microscopy in 1955 by Marvin Minsky (1927-2016) \cite{Minsky1988}.

One interferometric technique that would thrive throughout the 1960-70s was Interferometric Reflection Microscopy (IRM). This simple technique, which in fact does not require an explicit interferometer, relies on the sample cover-slip to provide the reference beam. The rise of IRM is widely accredited to Adam Curtis (1934-2017), who investigated and estimated the \textit{nanometric} separation between glass and cell at sites of adhesion in living cultured fibroblasts \cite{CURTIS1964}. The interferometric nature of the image provided nanometer-level information on the cell-substrate distances - a resolution unmatched by fluorescent methodologies of the time, enabling researchers using IRM to pioneer investigations into cell adhesion \cite{Ploem1975a,Abercrombie1975,Izzard1976,Heath1978,Wehland1979} as well as associated cytoskeletal components \cite{Opas1984}. Johan Sebastiaan Ploem (born 1927) would later introduce instrumental refinements to better the sensitivity and performance of the technique under the name alternative name Reflection Contrast Microscopy (RCM) \cite{Ploem1975a}. Ultimately the failure to interpret quantitatively the interferometric content of the images, i.e. accounting for contributions from variations in refractive index or path length \cite{Ploem1975a, Bereiter-Hahn1979}, would lead to declining interest in the method. Nevertheless, the technique would undergo a new lease of life in the late 1980s under the helm of Erich Sackmann (born 1934) who sought to implement greater quantitative rigor, initially studying cell membrane elasticity \cite{Zilker1987} and later the interaction of synthetic vesicle membranes to their substrates \cite{Radler1993,Wiegand1998a}. 

Other interferometric methods would also find a revival in the 1980s, following the realization that analogue controls for camera gain and offset can be used to increase the dynamic range and contrast of microscopic images. This contrast enhancement facilitated real-time observations of structures that were previously too faint to see by the naked eye. Indeed, weak contrast variations composing an image are more easily detected by the video camera owing to its linear response to light intensity. While this fact was known since the invention of the video camera in the 1940s, the better resolution and historical success of photography cemented the latter as the dogma in microscopic documentation. It would not be until 1975 that this view began to change \cite{Dvorak1975}, and another six years before the startling discovery of \textit{analogue} contrast enhancement \cite{Allen1981a,Inoue1981}.

The improvement in analogue detection contrast immediately precipitated a new problem of how to remove unwanted background signals such as \textit{mottle} originating from uneven illumination or the presence of dust, etc. By \textit{digitizing} the analogue video signal, these problems could be tackled because the live video signal could undergo in-line digital processing including static image subtraction, averaging and differential subtraction \cite{Walter1981,Allen1983}. This technique was referred to as \textit{digital} contrast enhancement, but along with its analogue cousin, would collectively become known as \textit{video-enhanced microscopy} \cite{Salmon2003,Inoue1986} and often paired with DIC imaging. Following its introduction in the early 1980s, video-enhanced microscopy quickly led to a flurry of efforts to explore previously hidden aspects of cellular architecture and function, including the structure of microtubules and actin filaments, or transport of particles and vesicles on and around the cell \cite{Brady1982, Allen1985, Schnapp1985, Weiss1986, Herman1984}. 

Following these demonstrations, the video-enhanced microscopy of the late 1980s and early 1990 quickly advanced efforts to monitor intracellular dynamics of specific proteins in living cells through introduction of sub-hundred nanometer colloidal particles. Michael Sheetz and colleagues pioneered this microscopy under the moniker \textit{Nanovid-microscopy} (nanometer video-enhanced microscopy) and investigated \textit{in vitro} mobility assays on mictotubules, receptor mediated endocytosis and single protein diffusion in the plasma membrane of cells \cite{Brabander1986,Geerts1987a,Schnapp1988,DeBrabander1988a,Sheetz1989,Geerts1991a} - a body of work that serves as the foundation for single-particle tracking microscopy. In the latter half of the 1990s to early 2000s, Akihiro Kusumi would use this technique to further study single protein diffusion in the plasma membrane of cells \cite{Kusumi2005e,Kusumi2012}. 

It is worth noting that the challenge of detecting and analyzing micro- and sub-microscopic particulates was also confronted in the field of aerosol science, dating from the period of the 1970s. Whilst employing intensity-based detection, in 1982 Pettit and Peterson \cite{Pettit1982} introduced an interferometric scheme derived from the Jamin interferometer to detect the phase shift induced by microscopic particles. By using phase, a better estimate of the particle size was demonstrated along with an improved detection sensitivity. This scheme would be refined further by Batchelder and Taubenblatt in 1989 \cite{Batchelder1989a}. 

The 1980s also saw the field of nanoscience emerge out of the collective invention of advanced scanning probe techniques, such as scanning tunneling microscope (STM) in 1981, scanning near-field optical microscopy (SNOM) in 1984 \cite{Pohl1984}, and the atomic force microscope (AFM) in 1986 \cite{Binnig1986} as well as the discovery of new materials such as carbon nanotubes and fullerenes. Optical microscopy would similarly undergo great advances in performance following demonstration of the optical detection of single dye molecules in 1989 \cite{Moerner1989}. Complemented by the general political optimism at the end of the cold war, these developments launched a euphoria in accessing the \textit{nano-world} at the turn of the 1990s, as had been imagined by Richard Feynman almost thirty years prior \cite{Feynman1960}. 

With the emergence of SNOM, optical microscopy experienced a revolutionary era as it was now freed of the century-long bounds of the diffraction limit, and brought promise of the detection and spectroscopy of individual molecules in contexts ranging from materials science to chemistry, physics and biology. As in conventional light microscopy, the main contrast mechanism in SNOM is based on extinction of light, now exiting a scanning sub-wavelength aperture. However, the physical interaction at work in SNOM should be understood in the context of \textit{scattering} rather than reflection or transmission as considered in far-field imaging. In practice, SNOM suffered from very weak signals stemming from faint illumination out of a nanometric aperture and weak scattering from nanoscopic features on the sample surface. One of the efforts in tackling the limited throughput was put forth in 1994 by the group of Wickramasinghe, who demonstrated interferometric (homodyne) detection of the field scattered from a sharp solid tip (apertureless SNOM) \cite{Zenhausern1994}. Interpretation of the SNOM signal in this mode and similar configurations presented a great challenge \cite{Sandoghdar1997}, which hampered adoption of this technique until they were overcome a few years later \cite{Hillenbrand2001,Taubner2003,Amenabar2013b}.

In this nascent era of nanoscience, reappraisal of the optical functionality of gold and silver nanoparticles and thin films, under the modern label of \textit{Plasmonics}, led a new generation of scientists to explore the detection and spectroscopy of individual metallic nanoparticles. The first spectroscopy reported from single gold nanoparticles appeared in 1998 using SNOM \cite{Klar1998}. Detection of individual gold nanoparticles was soon pursued by a number of methods, including conventional dark-field \cite{Schultz2000} and total internal reflection dark-field microscopy \cite{Sonnichsen2000a}. A particularly promising impetus was the biocompatibility of gold and its indefinite photostability, which prompted the use of colloidal nanoparticles as optical immunolabels for biology \cite{Schultz2000}. Interestingly, while just a few years earlier the common wisdom would have been that Rayleigh scattering is too weak for the detection of individual nanoparticles, these techniques equally found applications in the detection of other single nanoparticles such as carbon nanotubes \cite{Sfeir2004}. 

By the end of the twentieth century, the achievements in detecting single nanoparticles with diameter on the order of 50\,nm marked a milestone in detection sensitivity \cite{Klar1998,Schultz2000,Sonnichsen2000a}. However, most applications, especially those concerning biology, wished to work with much smaller particles, e.g. 5\,nm in diameter, corresponding to a million-fold reduction in signal strength for dark field microscopy. To address this huge challenge, other techniques such as lock-in enhanced photothermal detection \cite{Boyer2002b,Berciaud2004}, interferometric scattering (iSCAT) \cite{Lindfors2004} and lock-in enhanced transmission \cite{Arbouet2004} were soon introduced. The success of these measurements showed that sensitive optical detection of nano-objects would indeed be possible through non-fluorescent means. In the decade that followed right up to the present day we see a broad array of research efforts, drawn from numerous communities, exploring interferometric detection of single nano-objects such as viruses, DNA, microtubules, exosomes, and proteins \cite{Simmert2017,Mahamdeh2018,Daaboul2010,Avci2015,Yurt2011,Matlock2017,Scherr2016,Daaboul2016,Daaboul2017,Sevenler2017a,Sevenler2018,Wang2010,Yu2014a,Yang2018,Shan2012,Koch2018,Kandel2017}. Interestingly, interferometric microscopies are also flourishing in the general context of label-free imaging of cells and membranes even if nanoparticles are not at the center of attention \cite{Schilling2004b,Limozin2009,Matsuzaki2014a,Matsuzaki2016,Dillard2016,Dejardin2018,Junger2016,Junger2018,Koch2018,Contreras-Naranjo2013b,Simmert2017,Mahamdeh2018,Chiu2012,Matsuzaki2014b,Redding2014,Sencan2016a,Schain2014,Nguyen2017}. The underlying physics of these methods remains the same although a plethora of acronyms such as interference reflectance imaging sensing (IRIS) \cite{Daaboul2010}, rotating coherent scattering (ROCS) \cite{Junger2016}, interference plasmonic imaging (iPM) \cite{Yang2018}, coherent bright-field imaging (COBRI) \cite{Huang2017a}, stroboscopic interference scattering imaging (stroboSCAT) \cite{Delor2018}, interferometric scattering mass spectrometry (iSCAMS) \cite{Young2018} are on the rise. In what follows, we bring all these techniques under the umbrella of iSCAT, emphasizing the two central concepts and mechanisms of \textit{interference} and \textit{scattering} as the basis for recording the extinction signal (nano-shadow) generated by nanoparticles.

\section{Interferometric scattering microscopy (iSCAT)} 
\subsection{Foundations}
\label{theory}

\begin{figure}[thtp!]
   \centering
   \includegraphics[scale=0.5]{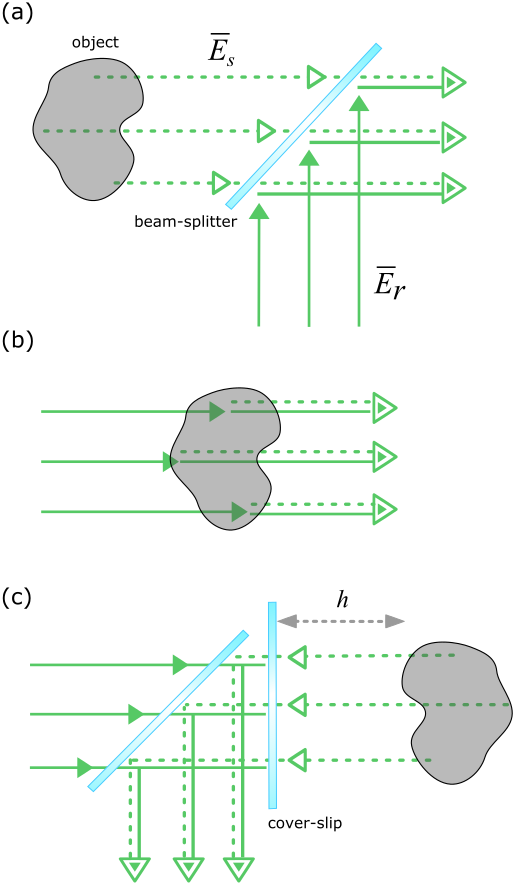}
   \caption{Different realizations of interferometric detection. (a) Principle of interferometry, where the signal of interest, here the scattered field $\overline{E}_{s}$ is superposed with a coherent reference $\overline{E}_{r}$. (b) Conventional bright-field imaging also consists of the superposing of reference (illumination) and scattered field, thus also interferometric in nature. (c) An alternative back-reflection scheme for interferometric detection.}
   \label{fig:interfer-microscopy-schemes}
\end{figure}

The principle concept in interferometric microscopy is to superpose a reference light beam with the response of the sample, as illustrated in Figure\,\ref{fig:interfer-microscopy-schemes}a. Let us consider a field $\overline{E}_{s}\,=\,E_{s}e^{i\phi_{s}}$ scattered from the object. The signal on the detector reads: 

\begin{equation}
I_{det} \propto |\overline{E}_{r} + \overline{E}_{s} |^{2} = I_{r} + I_{s} + 2E_{r}E_{s} \cos\phi\,,
\label{eqn_interf}
\end{equation}
where $\overline{E}_{r}=\,E_{r}e^{i\phi_{r}}$ denotes the complex electric field of the reference arm. The three resulting components can be respectively identified as the contribution of the reference field ($I_{r}=|\overline{E}_{r}|^{2}$), the pure scattering recorded from the object ($I_{s}=|\overline{E}_{s}|^{2}$) and finally the cross-term ($2E_{r}E_{s} \cos\phi$), wherein $\phi\,=\,\phi_{r}-\phi_{s}$. In general, this phase contains a component describing the  sinusoidal modulation along the propagation path, a Gouy phase stemming from variations of wavevectors and the scattering phase imposed by the material properties of the object \cite{Siegman1986,Pang2011}.

As is common in holography, one can realize various iSCAT illumination and detection schemes, which are all essentially described by eqn.\,(\ref{eqn_interf}). The simplest, oldest and perhaps the most subtle version is shown in Fig.\,\ref{fig:interfer-microscopy-schemes}b, corresponding to conventional bright-field imaging. Although the intuitive explanation of this imaging modality is based on the concept of absorption and shadow, its mathematical essence is well described by eqn.\,(\ref{eqn_interf}) if one only replaces $\overline{E}_{r}$ with an illuminating field $\overline{E}_{i}\,=\,E_{i}e^{i\phi_{i}}$ incident upon the object. Here, it is important to remember that as formulated by the optical theorem, extinction (loss of light, i.e., shadow) is determined by the cross-term in eqn.\,(\ref{eqn_interf}) and can be expressed as the sum of absorption and scattering \cite{Bohren-Huffman}. The latter two stem from imaginary and real parts of the complex extinction coefficient of the object, which are also implicitly encoded in $\phi_{s}$. As illustrated by the dashed and solid arrows in Fig.\,\ref{fig:interfer-microscopy-schemes}b, propagation phases are the same for $\overline{E}_{r}$ and $\overline{E}_{s}$ for each nanoscopic constituent of the object such that this imaging modality does not reveal any path dependence. To that end, the coherence length of the light source is not a parameter of concern. Another point of view that supports this picture is entailed in Abbe's theory on image formation and in Fourier optics, where an image results from the diffraction (and thus interference) of light from the sample. 

An interesting simple extension of the scheme in Fig.\,\ref{fig:interfer-microscopy-schemes}b is shown in Fig.\,\ref{fig:interfer-microscopy-schemes}c, where a partially reflective surface is placed in the illumination path. In this configuration, a portion of the illumination is reflected and interferes with the light that is back-scattered by the object. Thus, in this case we can write $\overline{E}_{r}\,=\,r\overline{E}_{i}$, where $r$ denotes the complex field reflectivity. In this instance, the detected signal depends sensitively on the displacement $h$ above the surface or index modulations of the object. 

Schemes such as phase contrast, differential interference contrast, reflection interference, or Mirau interference microscopy can all be described with the same underlying physics of eqn.\,(\ref{eqn_interf}) if one accounts for specific assignments of $\overline{E}_{r}$ and $\overline{E}_{s}$. In the conventional realizations of these microscopies, the emphasis has been on visualization of edge contours or refractive index modulations in super-wavelength objects or features. It turns out, however, that interference microscopy can be even more important for the detection of very small nanoparticles and single molecules, which is particularly desirable within the context of nanoscience.

Let us take a spheroid with semi-axes $a_1$, $a_2$, $a_3$ much smaller than the wavelength of light as a model nanoparticle. The response of this particle to light can be formulated as $\overline{E}_{s}\propto\,\alpha\overline{E}_{r}$ where
\begin{equation}
\alpha_i = \epsilon_{0}V\Bigg(\frac{\epsilon_{s}-\epsilon_{m}}{\epsilon_{m} + L_{i}(\epsilon_{s}-\epsilon_{m})} \Bigg)
 \label{eqn:polarisabilty}
\end{equation}
denotes the complex particle polarizability along the semi-axis $a_{i}$, and $k$ signifies the wavenumber \cite{Bohren-Huffman}. The quantity $V$ is the volume of the particle, and $L_{i}$ is the depolarization factor along $a_{i}$. For a sphere, $a_{1}=a_{2}\,=\,a_{3}$ and $L_{i}=1/3$. Complex parameters $\epsilon_{s}$ and $\epsilon_{m}$ are the dielectric functions of the scatterer and the embedding medium respectively, and $\epsilon_{0}$ is the dielectric constant. The denominator in this expression can reach a minimum for metals such as silver and gold, thus enhancing their scattering response, but this enhancement only amounts to about one order of magnitude for realistic materials in the visible domain. Thus, the main factor responsible for the magnitude of $\alpha_{i}$ is the particle volume $V$. The dependence on the volume results in the scattering field scaling as the third power of the particle dimension, and hence the scattering intensity ($I_{s}\,=\,|\overline{E}_{s}|^2$) drops with the \textit{sixth} power of the particle size. 

When the particle is sufficiently small, $I_{s}$ becomes much weaker than the cross-term $2E_{r}E_{s} \cos\phi$ in eqn.\,(\ref{eqn_interf}) which is linearly proportional to the scattered field. As a result, interferometric detection of scattering seems more favorable over dark-field schemes. Indeed, eqn.\,(\ref{eqn_interf}) highlights the challenge in dark-field microscopy of detecting small particles: the difference between the signal of a 50\,nm nanoparticle and one with 5\,nm diameter is a factor of \textit{one million}! We also remark that eqn.\,(\ref{eqn_interf}) describes both homodyne and heterodyne detection schemes \cite{Haus2000}, where the particle field is amplified by the larger field of the reference ${E}_{r}$ in the cross-term. We present a more differentiated discussion on the comparison between the sensitivities of dark-field and iSCAT detection later in this section.

The first report of iSCAT microscopy and spectroscopy of \textit{single} nanoparticles appeared in 2004 \cite{Lindfors2004} using the configuration shown in Fig.\,\ref{fig:interfer-microscopy-schemes}c (see section \ref{showcase} for an additional historical anecdote). More precisely, a supercontinuum laser beam was focused on a glass substrate supporting gold nanoparticles (GNPs) as small as 5\,nm covered by immersion oil. This was quite an impressive step towards optical detection of small nanoparticles since conventional methods such as dark-field microscopy were not able to reach this limit. The only other technique with appreciable sensitivity was reported just two years earlier based on photothermal detection \cite{Boyer2002b}. In this technique, one heats the GNP through its enhanced absorption at the plasmon resonance and detects the heat-induced change of refractive index in its vicinity. Interestingly, however, this latter decisive step is also achieved via interferometry using a second laser beam.

In the following years, iSCAT was extended in our laboratory to different illumination and detection conditions \cite{Jacobsen2006a,Kukura2009} and used to detect single unlabeled viruses \cite{Ewers2007,Kukura2009a}, semiconductor quantum dots \cite{Kukura2009}, lipid vesicles \cite{Krishnan2010,Spindler2016} and unlabeled proteins \cite{Piliarik2014}. In more recent years, several other groups have also successfully applied different illumination/detection variants of iSCAT to detect single proteins \cite{OrtegaArroyo2014f,Arroyo2016,Young2018}, single viruses \cite{Huang2017a,Daaboul2017}, lipids \cite{Andrecka2013, DeWit2015} and other nanoparticles \cite{Daaboul2016}, and even charge carriers \cite{Delor2018}.

Neglecting the scattering intensity ($I_{s}$) which is vanishingly weak for very small particles, the iSCAT signal of interest, namely the interferometric cross-term, can be obtained by subtracting the reference intensity that acts as a background from the detected intensity, $I_{det} - I_{r} \approx 2E_{s}E_{r}\cos\phi$. Thus, the contrast obtained when comparing images with and without a nanoparticle becomes:

\begin{equation}
c = \frac{2E_{s}E_{r}\cos\phi}{I_{r}}=2\frac{E_{s}}{E_{r}}\cos\phi\,.
\label{eqn_contrast}
\end{equation}

The final expression in eqn.\,(\ref{eqn_contrast}) might prompt one to conclude that one can reach a better sensitivity through minimization of ${E_{r}}$ in the denominator. Indeed, one can devise an optical scheme, where ${E_{r}}$ is freely adjustable, e.g. by implementing a separate reference arm (see, e.g. scheme of Fig.\,\ref{fig:interfer-microscopy-schemes}a). However, the resulting increase in contrast is at the cost of a lower overall signal which leads to a smaller signal-to-noise ratio (SNR) in shot-noise-limited detection. Furthermore, a separate arm comes compromises sensitivity owing to the introduction of mechanical instabilities. We discuss the issue of SNR and other instrumental considerations in more detail below.  

Before we begin the discussion of experimental issues, we take a moment to remark on the fundamental efficiency of interferometric detection of small objects, down to single atoms and molecules. When an ideal two-level atom is illuminated with monochromatic light at its resonance frequency, it undergoes a transition from the ground to the excited state. This process can lead to absorption and successive emission of light. The resulting fluorescence is incoherent because it follows a spontaneous process, thus not respecting the phase of the illumination field. However, in the weak excitation regime, the interaction of the incoming light with an atom can also be described by Rayleigh scattering \cite{Loudon}. It follows then that the interaction of the incoming light and the atom can equally be described in the same way as in eqn.\,(\ref{eqn_interf}), that is, via interference. Considering that the extinction cross-section of an unperturbed atom can be as large as $3\lambda^2/2\pi$ where $\lambda$ is its transition wavelength, and that light can be focused down to the diffraction limit in the order of $(\lambda/2)^2$, one ought to expect a single atom to be able of casting a dark shadow on a laser beam. Indeed, one can rigorously show that a single atom can extinguish a laser beam in entirety \cite{Zumofen2008}. 

The most fundamental requirement for reaching full extinction is spatial mode matching: the wavefronts of the incident beam should match the dipolar emission pattern of the atom \cite{Zumofen2008}. An alternative approach is to modify the radiation pattern of the atom by coupling it to an appropriate antenna \cite{Lee2011}. In fact, a glass-water interface is known to act as a primitive planar antenna that modifies the radiation pattern of an oscillating dipole into one that is enhanced about the critical angle \cite{Lukosz1979}. Optimization of the iSCAT contrast in recent efforts, where part of the reflected beam is blocked with a pinhole in the back focal plane, can be understood as the result of a better mode matching to such an emission pattern \cite{Liebel2017,Cole2017a,Avci2017b}. A particularly compact antenna consists of a subwavelength waveguide (nanoguide), such as a thinned glass fiber \cite{Faez2014}, whereby the emission of an atom is efficiently coupled to the nanoguide mode, thus efficiently interfering with the field propagating within. Given the central role of spatial modes in this picture, these considerations are also of immediate relevance for iSCAT detection of nanoparticles, which radiate with a dipolar pattern. To our knowledge, this feature has not yet been explored although the nanoguide arrangement has been demonstrated to achieve a high-performance dark-field microscopy \cite{Faez2015a}.

\subsection{Detection sensitivity and signal-to-noise ratio (SNR)}
Thus far, iSCAT has been successfully employed to detect individual unlabeled proteins as small as 50\,kDa. An important question that arises is whether there exists a fundamental limit in detection sensitivity, and what might be the accompanying technical challenges. In what follows below, we give an overview of some of the pertinent issues.

The central restriction in any sensitive measurement is signal fluctuation. In their absence, arbitrarily small signals may be identified even on very large backgrounds. While slow fluctuations such as thermal or mechanical drifts can be directly or indirectly accounted for, fast random variations, which we call "noise" pose a serious challenge. Some of the main sources of noise in an iSCAT measurement are:

\underline{Laser intensity noise}: Even the best lasers have instrumental power fluctuations and beam instabilities. It is indeed not easy to have a freely-running laser with power stability better than about $10^{-3}$. To detect iSCAT contrasts beyond this, one has to account for laser intensity fluctuations through referencing or normalization. In confocal imaging, a balanced photodiode pair employing common-mode rejection can be as good as 1$\times 10^{-7}$ \cite{Celebrano2011a}. In wide-field camera-based detection, one can normalize the total power recorded within each frame to similarly reject intra-frame fluctuations. As the contrast signal is the ratio of two fields sharing the same laser noise, they naturally self-reference.

If one manages to deal with the instrumental laser intensity noise, one is still confronted with the fundamental limit of shot noise, the noise associated with the fact that the number $N$ of the photons in a laser beam varies according to a Poisson distribution, i.e. as $\sqrt{N}$ for large $N$ \cite{Loudon}. As a rule of thumb, the shot-noise-limited SNR improves by $\sqrt{N}$ as $N$ is increased.

\underline{Detector background noise}: Any electronic device has an intrinsic noise, for example, stemming from thermally generated electrons in the detector (referred to as dark noise) or errors introduced in the voltage reading circuitry. Modern cameras and photodiodes can be extremely quiet, and since iSCAT is typically performed on a high background level of the reference field $E_{r}$, detector dark noise is not a major concern. Nevertheless, depending on the experimental arrangement, it might become a limiting factor. 

\underline{Dynamic range and analogue-to-digital conversion noise}: Realistic detectors have a limited working range on both sides of small and large signals. The limit for the largest signals is given by detector nonlinearities and saturation effects, while the lower limit often has to do with the fundamental sensitivity of the particular detector technology. The ratio between the largest and smallest signal values defines the dynamic range, and for imaging cameras the read noise level is often taken as the smallest signal quantity which is detectable. From the sensor dynamic range, an appropriate bit depth is selected for analogue-to-digital conversion. For example, an image rendered into 12-bit imposes a read-out resolution of 1 in 4096, i.e. 2$\times 10^{-4}$. 

\underline{Mechanical stability}: Although iSCAT is an interferometric method, it can be extremely robust against mechanical instabilities if the reference and the scattering beams share paths (see Fig. \ref{fig:interfer-microscopy-schemes}). Nevertheless, lateral vibrations might cause problems since even a few nanometers of motion could translate into fluctuations of the contrast when the background is subtracted (see section below) \cite{Gemeinhardt2018}.

\subsection{Background Removal} 
Fluorescence detection exploits highly efficient spectral filtering to eliminate spurious background caused by the illumination or unwanted fluorescence. Similarly, dark-field microscopy, including variants using total internal reflection, employ spatial instead of spectral filtering to detect Rayleigh scattering. In iSCAT, however, one does not exclude the background but instead measures on an intense reference beam, just as in resonant extinction measurement of a quantum emitter \cite{Wrigge2008}. 

\begin{figure}[!htb]
    \centering
   \includegraphics[scale=.65]{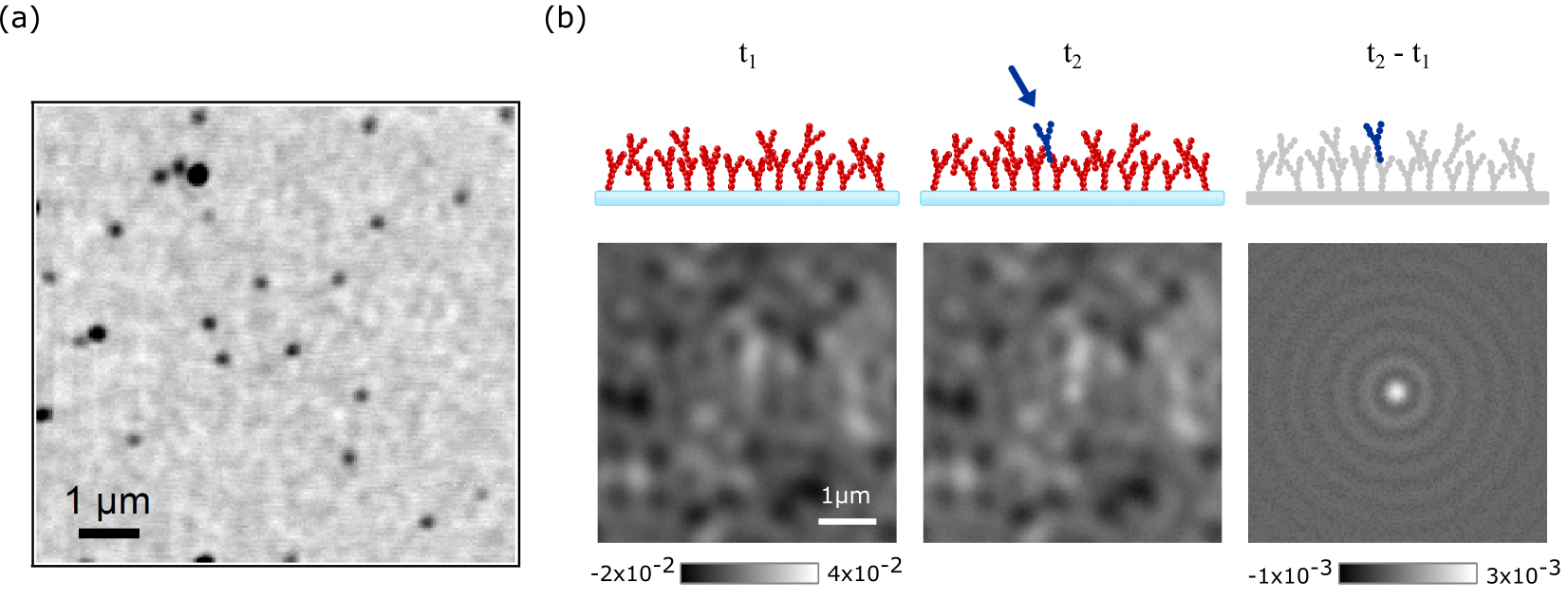}
    \caption{(a) Substrate roughness introduces modulations in the background of the image in accompaniment to imaging of gold nanoparticles of size 10\,nm \cite{Jacobsen2006a}. Reproduced with permission from the Optical Society of America. (b) Background subtraction through differential imaging. In wanting to image the arriving protein shown in blue, one subtracts the image of the substrate at a time before ($t_{1}$) arrival, and one after ($t_{2}$). The difference isolates the presence of the target protein. }
    \label{fig:Jacobsen}
\end{figure}

As mentioned earlier, this would not pose any problem if one could subtract a constant background level from the measured signal, even if the signal was to be arbitrarily small. Figure \ref{fig:Jacobsen}a, which displays an iSCAT image of a cover-slip containing 10\,nm GNPs embedded in water, shows that in practice, one is confronted with lateral background modulations that make it difficult to identify the gold particles. These background features, however, should not be attributed to noise since they are fully reproducible: because iSCAT is extremely sensitive to slightest changes in the optical path - down to the level of single small proteins - the background contains a high degree of a speckle-like pattern caused by a slight inhomogeneity of the refractive index or topography.  

Background components that do not originate from the sample and its environment can be eliminated by measures such as lock-in-type detection. For example, wavefront inhomogeneities in wide-field illumination can be removed by mechanically modulating the sample \cite{Jacobsen2006a, Piliarik2014}. The most effective procedure, however, would involve modulations of properties that are specific to the nanoparticle of interest. For example, the wavelength dependence of plasmon spectra were used to separate the signal of GNPs from their dielectric background including microtubules \cite{Jacobsen2006a}. 

A very powerful method for eliminating the background becomes available in dynamic studies, where the particle of interest appears on the detection scene at a given time or moves within it. In this case, differential treatment of consecutive images can eliminate the static part of the sample \cite{Kukura2009,Piliarik2014}, illustrated in Figure\,\ref{fig:Jacobsen}b. This can be achieved by an assortment of methods, e.g. subtraction of a temporal median intensity \cite{Hsieh2014}, subtraction through an iterative-estimation algorithm \cite{Cheng2017b} or employing rolling-window averaging across stacks of frames \cite{Young2018}. The method chosen should be based on the problem and equipment at hand regarding image acquisition speed and also the speed at which the nanoparticle moves. The situation becomes more challenging in the presence of a background with a fluctuating spatio-temporal dynamics, e.g. speckle features from live biological specimens. Nevertheless, more advanced computational tools can be employed for analyzing the obtained images as recently demonstrated for tracking GNP-labeled transmembrane proteins in live cells \cite{Taylor2018}.

We remark that the background subtraction methods discussed above could also be equally well applied to dark-field microscopy \cite{Weigel2014b}. Thus, one might wonder whether dark-field microscopy might not match the performance of iSCAT microscopy. Indeed, one can show that the shot-noise-limited advantage of interferometric (homodyne) over dark-field (scattering intensity) detection is only a factor of two. The practical implementation of ultrasensitive dark-field, however, is nontrivial because the very small dark-field signal proportional to the sixth power of the particle size (see eqn.\,(\ref{eqn_interf})) puts higher demands on the detector technology.

\subsection{Long measurements: indefinite photostability} 
One of the key advantages of detecting scattering over fluorescence is that the former does not suffer from photobleaching. Whether one uses the inherent scattering of a bioparticle such as a virus or if one detects a GNP label, the scattering signal does not degrade over time. Thus, very long - in principle indefinite - measurements become possible. Such experiments might encounter technical difficulties such as the particle moving out of the field of view, but these can be easily overcome with more sophisticated instrumentation.

\subsection{Fast measurements: no saturation} 
Another favorable feature of scattering contrast as compared to fluorescence is lack of saturation. A fluorophore is a quantum mechanical system with an inherent anharmonicity, implying that only one photon at a time can be absorbed. The fluorescence lifetime of the excited state places a limit on how fast the photon can be emitted, imposing a bottleneck for the rate at which the fluorophore can radiate, and thus a limit to how fast one can image. A nanoparticle behaves like a classical oscillating dipole which does not suffer from saturation: the stronger the illumination, the higher the rate of scattering. Indeed, we have demonstrated iSCAT imaging speeds up to about 1\,MHz \cite{Spindler2016}. The immediate limitation is currently a technological matter of availability of suitable cameras. A more fundamental limit is introduced when concerned over photodamage of the sample when using very strong illuminations since every realistic substance also absorbs light at every wavelength, even if very weakly. When performing iSCAT on biological samples, illumination intensities on the order of 0.001-0.1\,mW$\mu$m$^{-2}$ have been reported for membranes \cite{Spindler2016,Taylor2018}, with powers as high as 5\,mW$\mu$m$^{-2}$ \cite{Spindler2016} for the fastest MHz imaging rates.

\subsection{Exquisite lateral and axial resolution} 
The resolution attainable in iSCAT imaging is diffraction-limited. However, when investigating single nanoparticles, it is often its location and trajectory that is of interest. These can be obtained in an analogous fashion to fluorescence localization microscopy and particle tracking \cite{Deschout2014a}. Lack of photobleaching and saturation provides strong iSCAT signals which in turn yield a higher localization precision within a shorter observation time than is achievable in fluorescence imaging.

\begin{figure}[!htb]
    \centering
   \includegraphics[scale=.65]{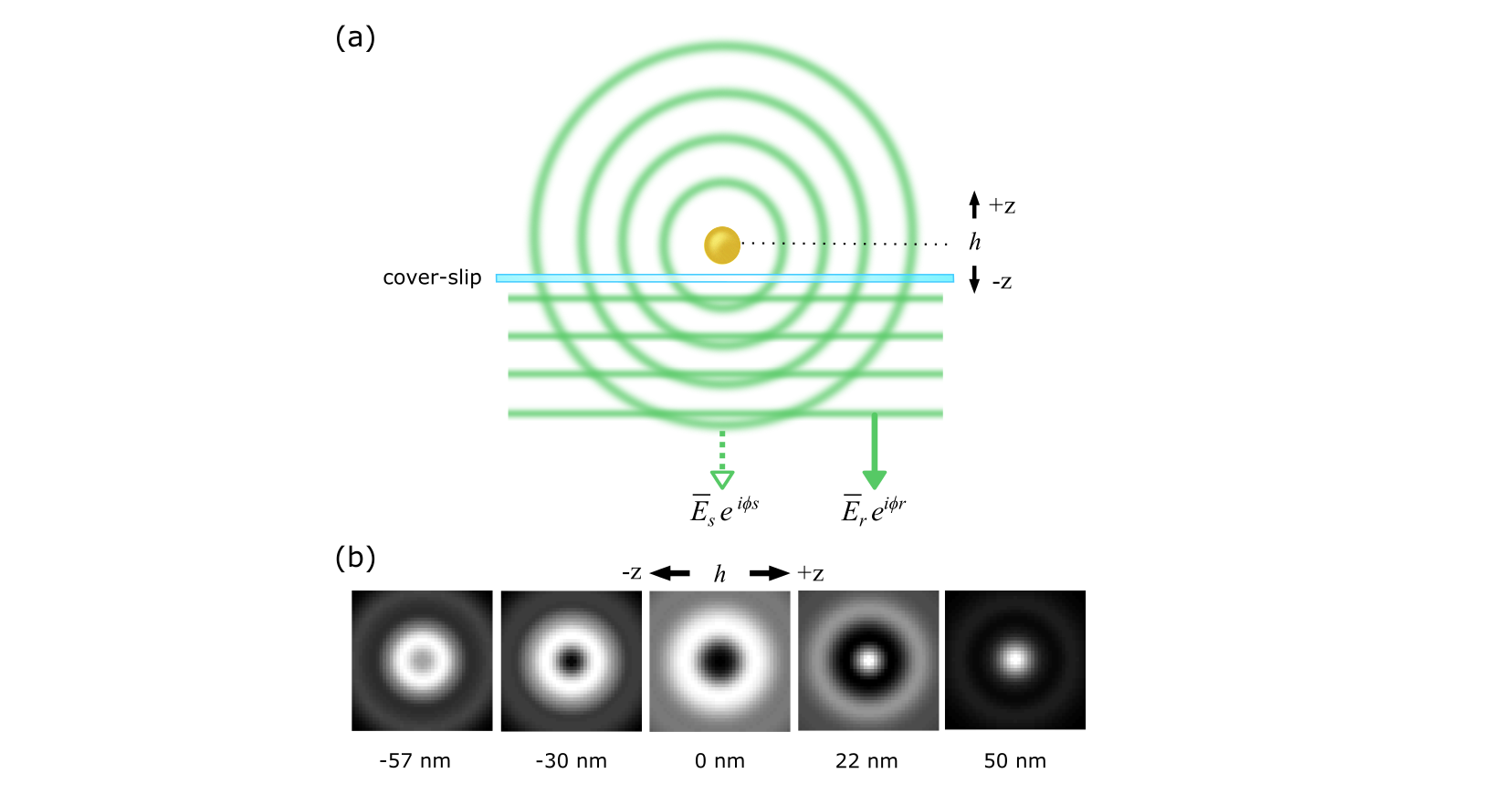}
    \caption{(a) Schematic view of the scattering specimen upon the cover-slip in iSCAT imaging. (b) The interferometric point spread function of the scatter in reflection wide-field mode possesses a unique signature of ring radii and contrasts which vary as a function of the axial position $\textit{h}$ of the scatterer above the reflecting cover-slip.}
    \label{fig:psfrings}
\end{figure}

The point-spread function (PSF) in iSCAT can take on different forms, depending on the illumination and detection modes. A particularly useful situation is encountered in wide-field illumination where plane waves and spherical waves interfere on the camera, resulting in many rings around the main PSF spot (see Fig.\,\ref{fig:psfrings}) \cite{Avci2016,Sevenler2017a,Taylor2018}. For most lateral tracking applications a Gaussian fit to the central spot is sufficient although the radial symmetry of the overall PSF can be very helpful in identifying small signals on a complicated feature-rich background. For example, GNPs have been tracked with spatial resolution of 2\,nm within 10\,$\mu$s on a live cell membrane \cite{Taylor2018}.

\begin{figure}[!htbp]
    \centering
   \includegraphics[scale=.65]{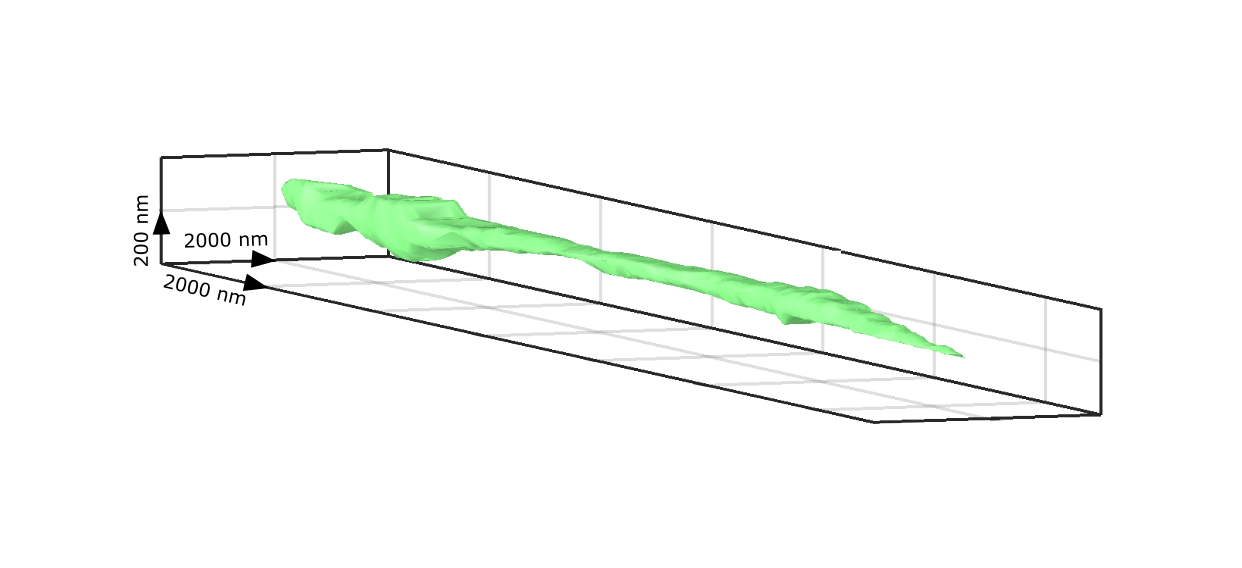}
    \caption{A reconstruction of the upper surface of a cellular filopodium, rendered from interpolation of approximately 800,000 three-dimensional trajectory positions as a GNP traverses the filopodium surface.}
    \label{fig:Intracellular-filament}
\end{figure}

A major asset of iSCAT imaging stems from its inherent interferometric nature, which makes the signal very sensitive to phase $\phi$ in eqn.\,(\ref{eqn_interf}). Consequently, axial displacements of a nanoparticle in the order of nanometers can be put into evidence. The first accounts of this nano-holographic feature of iSCAT were presented in Refs.\,\cite{Jacobsen2006a} and \cite{Krishnan2010}. In these studies, however, one was limited to displacements below 100\,nm by the ambiguity that is due to the periodicity of the iSCAT signal. More recently, we have shown that a quantitative PSF analysis in wide-field iSCAT gives access to nanometer axial resolution over several micrometers \cite{Taylor2018,Gholami2019}. Figure\,\ref{fig:Intracellular-filament} presents an example of a three-dimensional surface map of an intercellular filament generated by the motion of a GNP that was bound to an epidermal growth factor receptor (EGFR). We note that the best axial resolution is attainable in the reflection mode iSCAT, where the scatterer can be located at a distance above the beam-splitting cover-slip, and thus accumulate a traveling phase. 

\subsection{Illumination and detection schemes}
\begin{figure}[!htb]
    \centering
   \includegraphics[scale=.65]{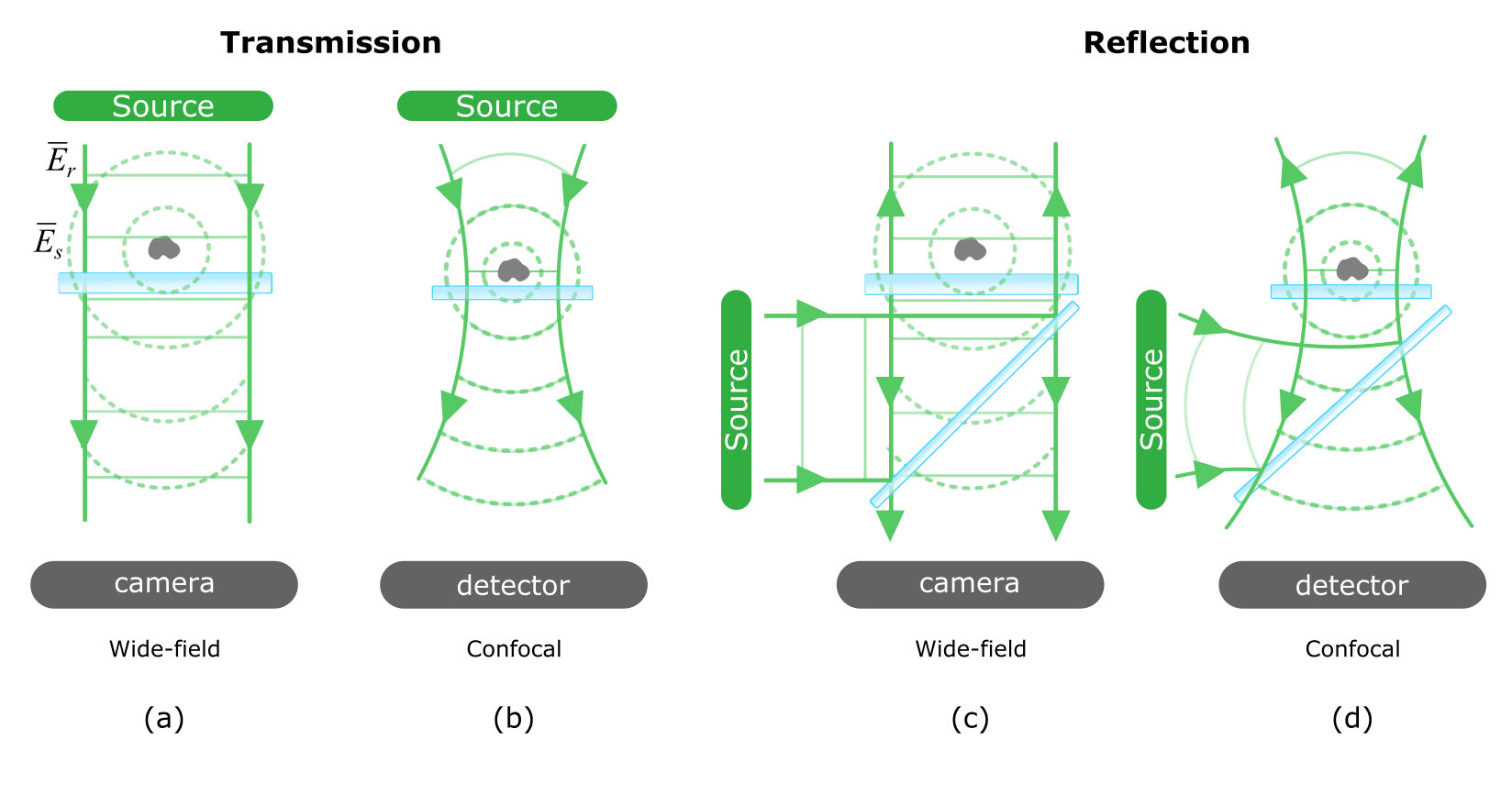}
    \caption{Modes of iSCAT microscopy, encompassing wide-field and confocal illumination and detection schemes performed in both the forward transmission or backward reflection geometry.}
    \label{fig:my_label}
\end{figure}

Interferometric detection of light scattering can be realized through various flexible permutations of illumination and detection schemes. The first efforts used scanning confocal point illumination and detection \cite{Lindfors2004} but this was soon extended to wide-field \cite{Jacobsen2006a} and fast beam-scanning \cite{Kukura2009a} illumination schemes in conjunction with camera-based detection. Of these, one additionally may place the detector in the forward direction or in the reverse (see Figure\,\ref{fig:my_label}), while keeping the inherent interferometric character of iSCAT.

In confocal point detection the reference consists of a focused Gaussian beam that is raster scanned across the sample. While this helps to discriminate unwanted scattering from the depth of the samples, a major hurdle is that temporal noise in the illumination light becomes translated into spatial noise in the image as each image pixel is acquired sequentially in time. Another noteworthy feature of a scanning mode is better mode-matching between the spherical scattered and reference waves, which renders the PSF without distinct ring features. 

For applications requiring fast imaging, wide-field illumination with camera-based detection is the most popular modality. By measuring in reflection, higher contrasts are obtained than for transmission because of the lower amplitude of the reference field in the former ($r\,=\,0.06$ for the reflected beam, see eqn.\,(\ref{eqn_contrast})). This assists greatly with visual inspection of the raw iSCAT image. In the case where one wishes to better the contrast for real-time inspection, phase-masks that attenuate the reference field have recently been introduced (recalling Lister's dark-field stop, or the Zernike phase plate) \cite{Cole2017a,Liebel2017,Avci2017b}. In the absence of computational real-time background removal, physical contrast enhancement can serve a useful function. 

The great advantage of reflection iSCAT is its sensitivity to the axial position of the nanoparticle. Transmission measurements, on the other hand, suffer less from the background since the disappearance of phase differences between $E_{r}$ and $E_{s}$ (see Fig.\,\ref{fig:interfer-microscopy-schemes}) minimizes speckle. This advantage was used in conjunction with index-matching for detecting single small organic molecules \cite{Kukura2010,Celebrano2011a}. What is important to realize, however, is that while most iSCAT measurements are performed with a laser, this is not a necessity if the distance between the nanoparticle and the place where the reference is picked up is not larger than the coherence length of the source. In other words, measurements involving nanoparticles very close to a cover glass can be just as well done using incoherent sources such as LEDs \cite{Simmert2017,Trueb2017, Daaboul2017, Sevenler2018}.

\section{iSCAT showcase}
\label{showcase}
\subsection{Detection and sensing of nanoparticles}
Since the debut of iSCAT in 2004, iSCAT and related techniques have revived interference and extinction microscopies in the context of detection and imaging of various nanoparticles. In this section, we provide a brief overview of some of the exciting application areas to which these methods have contributed.

\subsubsection{Gold Nanoparticles}
\begin{figure}[!htb]
\centering
   \includegraphics[scale=.65]{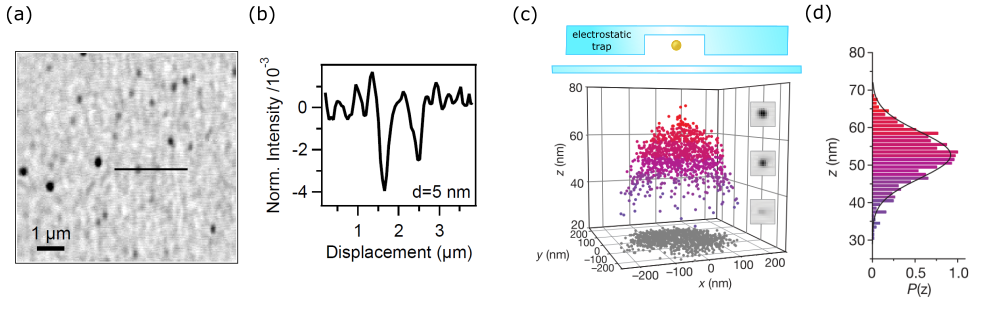}
    \caption{iSCAT microscopy on single nanoparticles of gold. (a) An early demonstration of iSCAT imaging of 5\,nm GNP at the glass-water interface. (b) The intensity profile corresponding to the cross-section marked in (a) \cite{Jacobsen2006a}. Reproduced with permission from the Optical Society of America. (c) Three-dimensional localizations of a 100\,nm gold particle within a 500\,nm electrostatic trap potential (illustrated above), revealing the potential landscape of the trap pocket. (d) Histogram of axial positions from which trap stiffness can be discerned \cite{Krishnan2010}. Reproduced with permission from the Nature Publishing Group.}
    \label{fig:sc-gold}
\end{figure}

Gold nanoparticles in the colloidal form have a wide range of applications due to some key properties such as biocompatibility, inertness, ease of fabrication and possibility of functionalization with different molecules. The recent developments of nano-optics, and in particular Plasmonics, have brought about a strong drive for studying and using these nanoparticles. Indeed, the first account of iSCAT was published on the direct far-field imaging of single GNPs, to size as small as 5\,nm. This was a formidable task at the time as in dark-field imaging one begins to struggle to visualize gold colloids smaller than 40\,nm. The GNPs, immobilized upon a cover-slip, were imaged when index-matched with oil \cite{Lindfors2004} and also under ambient conditions at the water-glass interface \cite{Jacobsen2006a} - see Fig.\,\ref{fig:sc-gold}a. The latter is especially important as it begins to match biological conditions wherein ultra-small GNPs serve as useful super-resolution probes (discussed later). The sensitivity limit on the size of the detectable GNPs in the early experiments was set by the speckle background that results from slightest variations in the refractive index and topography of the underlying glass substrate.  

To date, GNPs as small as 2\,nm in diameter \cite{Holanova2019}, smaller than a protein, have been imaged and localized to a precision of 8\,nm with short exposure times sufficient to function as a scattering label - cementing iSCAT as a powerful means to image and localize nanoscale colloids. At this point, it is perhaps interesting to mention that we had actually started exploiting iSCAT already in 2001 when investigating gold nanoparticles of diameter 100\,nm \cite{Kalkbrenner2001}. In those studies, we detected individual gold nanoparticles at the glass-air interface as dark spots in a scanning confocal reflection measurement and were puzzled that particles which scattered well would actually appear dark. Further studies then led to the appreciation of the role of interference and the advent of iSCAT \cite{Lindfors2004}.

Given that iSCAT imaging intrinsically contains quantitative phase information about the sample, proper interpretation of the interferometric scattering PSF allows one to attain material and morphological features of the colloid. For example, calibration of the spherical colloid size from the extinction contrast has been demonstrated \cite{Zhang2015}, as well as the orientation of anisotropic ellipsoidal nanorods through polarized detection \cite{Lee2018}. Moreover, the complete complex dielectric function of a single gold nanoparticle has been retrieved from colloids as small as 10\,nm by performing iSCAT with a supercontinuum light source in a DIC configuration \cite{Stoller2006}.

iSCAT tracking of GNPs has also been exploited to probe the three-dimensional landscape of electrostatic potential traps \cite{Krishnan2010,TaeKim2014} - shown in Fig.\,\ref{fig:sc-gold}c,d. In a similar fashion, the height occupation probability, which provides a description of the free-energy landscape within a microfludic slit channel, has been demonstrated for fast tracking of diffusing 60\,nm GNPs \cite{Fringes2016}. Geometry-induced electrostatic potentials at the end of a nanopipette have also been used for local manipulation of plasmonic antennas observed by iSCAT \cite{Tuna2017}. 

\subsubsection{Semi-conductor colloids and crystals}

The first half of the early 1990s showed that single dye molecules could be detected via fluorescence microscopy at room temperature \cite{Rigler2002}. The key for the success of these endeavors was efficient spectral filtering, efficient collection and sensitive low-noise detection of the emitted photons on a very low background. Once the dogma surrounding the difficulty of single-molecule detection was overcome, other fluorescent entities such as semiconductor quantum dots and diamond color centers were also detected in the same fashion. Considering that very few species fluoresce, however, this method found limited use, prompting scientists to search for alternative ways to detect nanoscopic amount of matter via extinction rather than fluorescence.  

\begin{figure}[!htb]
    \centering
   \includegraphics[scale=.7]{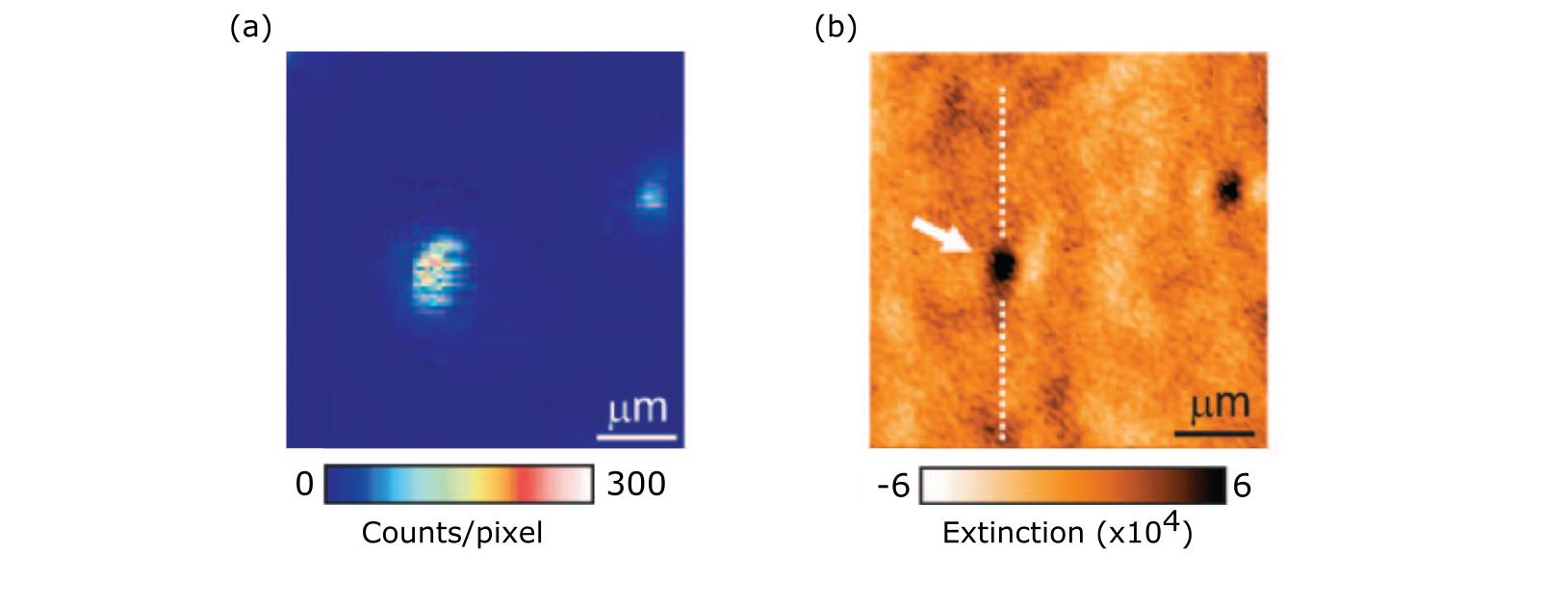}
    \caption{Imaging of single quantum dots through iSCAT. (a) A fluorescence image of two quantum dots on a substrate. (b) Transmission iSCAT image of the same region as (a) recorded in scanning mode, similarly revealing the interferometric extinction of the two nanocrystals identified in (a)  \cite{Kukura2009}. Reproduced with permission from the American Chemical Society.}
    \label{fig:sc-semicon}
\end{figure}

A single molecule may possess an extinction cross-section of $\sigma\,=\,10^{-16}$-$10^{-15}$\,cm$^{-2}$ whereas a diffraction-limited beam can be focused down to an area of $A\,=\,10^{-9}$\,cm$^{-2}$. Thus a simple estimate of $\sigma/A$ reveals the need for suppression of noise to the level of parts-per-million in order to overcome the challenge of detecting a single molecule in extinction. While several groups attempted absorption measurements in various configurations \cite{Hwang2006,Gaiduk2010,Chong2010a}, it was iSCAT that demonstrated direct room-temperature modulation-free extinction of single molecules and quantum dots \cite{Kukura2009,Kukura2010,Celebrano2011a}. 

The first important step in extending the sensitivity of iSCAT was taken with the detection of single semiconductor nanocrystals \cite{Kukura2009}, shown in Fig.\,\ref{fig:sc-semicon}. Core-shell semiconductor nanocrystals such as CdSe/ZnS have been employed in many areas of photonics, optoelectronics and bio-imaging. By varying the material and size of the core and shell, one can tune their optical properties such as emission wavelength and photostability over a wide range. 

The extinction cross-sections of the quantum dots under study were on the order of $10^{-15}\,$cm$^{-2}$. By using a thin sheet of mica as substrate, which can be locally atomically flat, the iSCAT background was reduced by about one order of magnitude. Furthermore, laser intensity fluctuations were accounted for by employing a second photodiode as a power reference. These measures turned out to be sufficient to detect different types of single core-shell colloidal dots. Simultaneous fluorescence and iSCAT measurements as well as iSCAT benchmarking with 10\,nm gold nanoparticles provided a robust evidence for the success of iSCAT in detecting extinction from individual quantum emitters at room temperature. A very interesting example of study that became accessible with such measurements is the investigation of the quantum dot even during dark periods of photoblinking \cite{Kukura2009}.

\begin{figure}[!htb]
   \includegraphics[scale=.65]{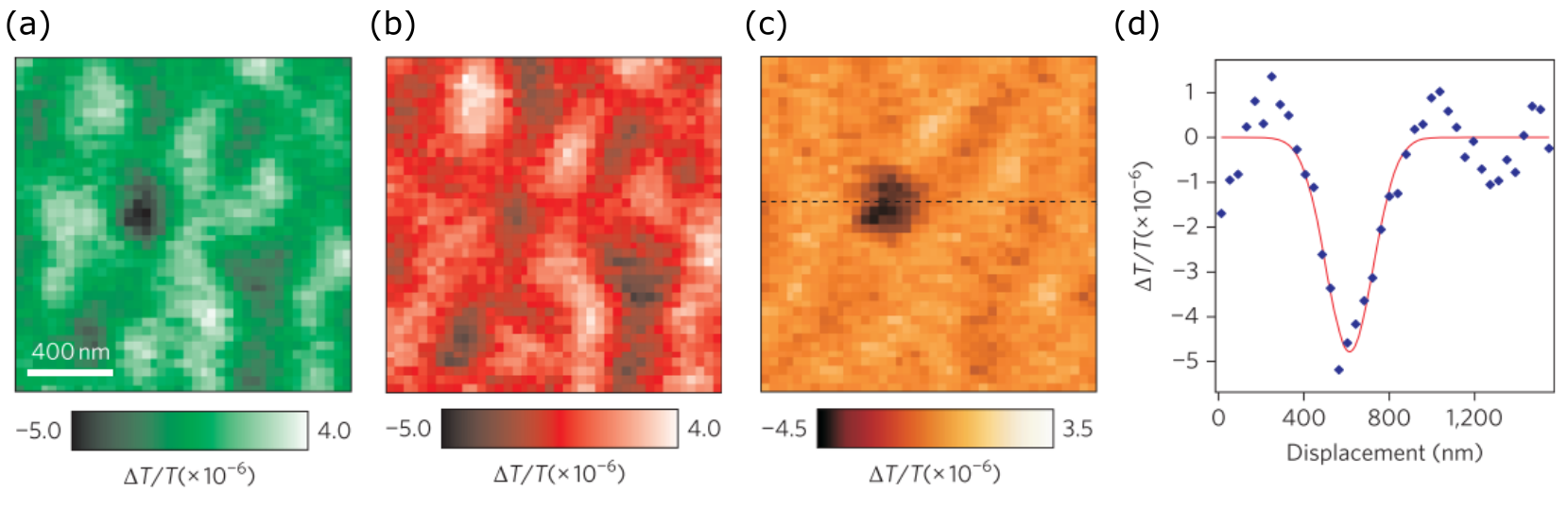}
    \centering
    \caption{Imaging of a single TDI molecule through confocal-scanned transmission iSCAT. Shown are the extinction raster-images of a single embedded dye molecule when illuminated (a) near resonance (633\,nm) and (b) off resonance (671\,nm). The difference between image (a) and (b) - shown in panel (c) clearly reveals the single molecule, and evidences wavelength-dependent detection of a molecule, a prerequisite for single molecule absorption spectroscopy. (d) Intensity profile through the cross-section marked in (c), revealing the extinction detection sensitivity required for single-molecule detection \cite{Celebrano2011a}. Reproduced with permission from Nature Publishing Group.}
    \label{fig:sc-sfm}
\end{figure}

To extend extinction experiments to single organic molecules, it was necessary to improve the SNR further owing to the smaller extinction cross-section of a molecule. Here, a better suppression of laser noise and of the background was required. To address the first issue, commercial balanced detectors were used \cite{Kukura2010} since self-assembled referencing solutions are accompanied by very small performance differences between the two detectors, e.g. as was used in the previous experiment on quantum dots, resulting in unwanted signal fluctuations. To improve the background issue, we chose to immerse the molecules in index matching oil and measure in transmission. These measures pushed the sensitivity beyond $1\,\times\,10^{-6}$ and led to the first successful direct detection of single-molecule absorption \cite{Kukura2010}. This work was progressed to the imaging of strongly quenched molecules and of molecules at different wavelengths \cite{Celebrano2011a}. By detecting a single molecule both on and off resonance, the way for single-molecule absorption spectroscopy was paved (see Fig.\,\ref{fig:sc-sfm}). We point out in passing that a transmission measurement of extinction is equivalent to an iSCAT measurement in reflection, which can be easily seen as a folded transmission experiment. A central feature of a reflection measurement is access to the traveling phase and thus to the axial position of the nanoparticle, which might complicate simple absorption spectroscopy. 
 
Let us now connect the underlying physics of the results presented above with those discussed in section \ref{theory}. When considering a two-level system in a non-ideal environment where the quantum transition is homogeneously broadened beyond the natural linewidth, the expression for the extinction cross-section is generalized to $\sigma=\frac{3\lambda^2}{2\pi}\times\frac{\gamma_{rad}}{\gamma_{tot}}$ where $\gamma_{rad}$ is the radiative linewidth and $\gamma_{tot}=\gamma_{rad}+\gamma_{nr}+\gamma_{deph}$ with $\gamma_{nr}$ and $\gamma_{deph}$ denoting the nonradiative and dephasing contributions to the linewidth, respectively. For nearly all emitters in the solid state, the quotient $\gamma_{rad}/\gamma_{tot}$ amounts to about $10^{-6}-10^{-5}$ at room temperature. It is now important to note that for most quenched systems this strong reduction is dominated by $\gamma_{deph}$, i.e. quenching is not the main factor. Hence, it follows that the extinction cross-section of a system that is quenched by as much as 1000 remains essentially the same. In other words, whether the system has a high quantum efficiency (given by $\frac{\gamma_{rad}}{\gamma_{rad}+\gamma_{nr}}$) or not is not decisive for extinction measurements. This is why quenched molecules can be detected in extinction but not in fluorescence.

Gold nanoparticles, quantum dots and dye molecules all have resonances, with their polarizability experiencing a maximum in a certain spectral range. While this enhances the extinction cross-section, and thus the iSCAT signal, it is not nearly as decisive as the influence of the particle size. A consequence is that even a dielectric nanoparticle can yield a large interferometric signal given it is of sufficient size. In what follows, we show that all viruses and a vast range of proteins easily satisfy this criterion, prompting efforts in their detection.

\subsubsection{Viruses}
\begin{figure}[!htb]
    \centering
   \includegraphics[scale=.65]{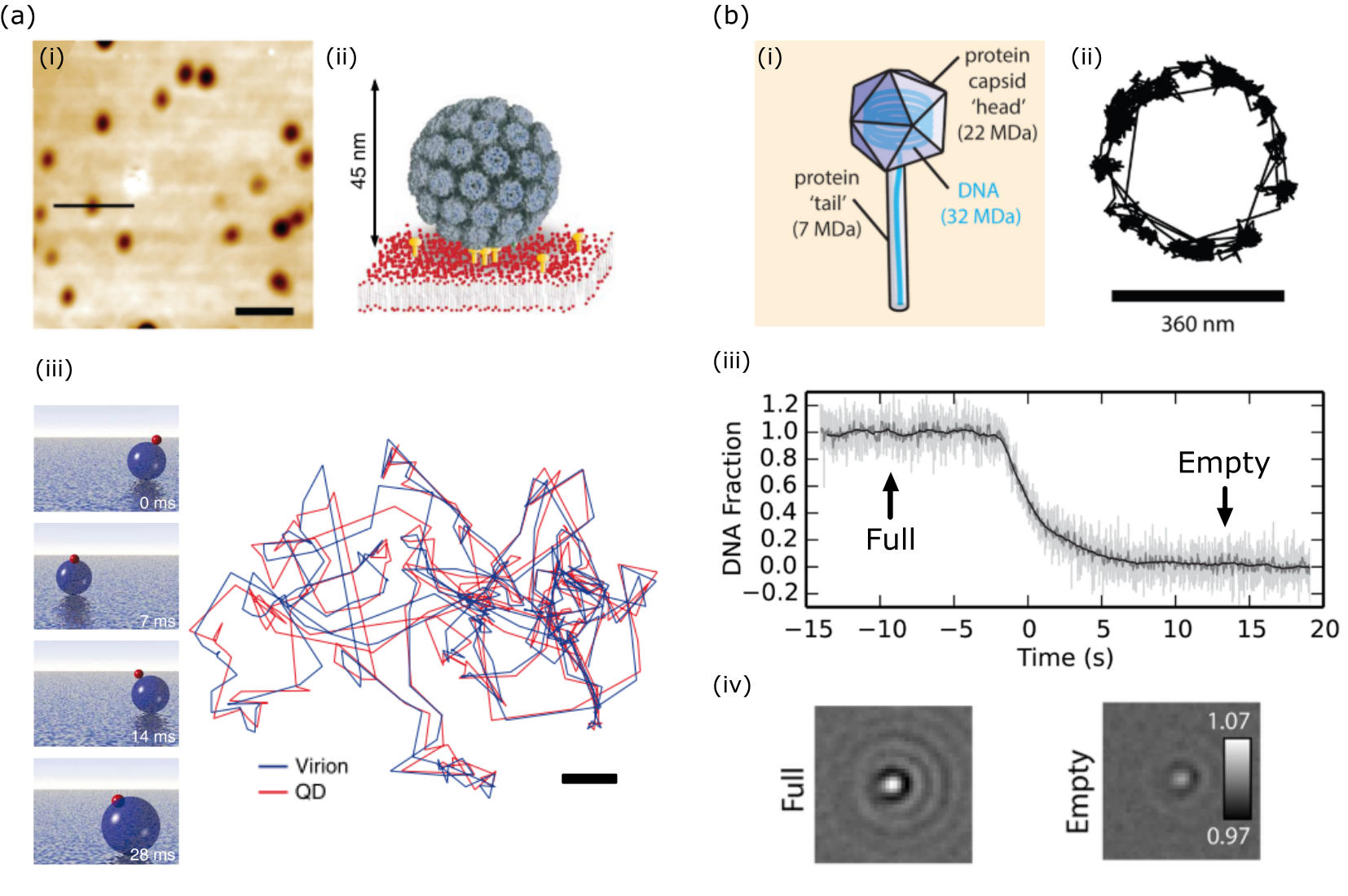}
    \caption{iSCAT microscopy on individual viruses. (a) The first demonstration of iSCAT detection of individual 45\,nm SV40 viruses: An iSCAT imaged when immobilised on glass, imaged via a confocal-reflection modality. Scale bar denotes 1\,$\mu$m and mean contrast of the viruses is 2\,\% \cite{Ewers2007}. Reproduced with permission from the American Chemical Society. When bound to a synthetic membrane, shown schematically in (ii) the trajectory of a single virus when interacting with a model membrane is possible to observe (iii). By simultaneous tracking of the virus and a quantum dot label on its surface, comparisons of both trajectories reveal the mobility to be composed of a rocking and tumbling motion. Scale bar denotes 100\,nm \cite{Kukura2009a}. Reproduced with permission from Nature Publishing Group. (b) Dynamic measurements of bacteriophages: (i) schematic of bacteriophage anatomy, (ii) when bound by the tail to the substrate, rotational motion of the bacteriophages can be tracked. (iii) Monitoring of DNA ejection out of the capsid head from the calibrated iSCAT contrast as a function of time. (iv) Images show corresponding video frames of a single bacteriophage while full and empty, each showing a differing contrast \cite{Goldfain2016}. Reproduced with permission from the American Chemical Society.}
    \label{fig:virus}
\end{figure}

Viruses play a crucial role in biology, whether causing harmful diseases or performing an integral symbiotic function within living systems \cite{Robinson2014} or even serving as novel disease treatment vectors \cite{Mietzsch2017}. Detection of viruses, as with most other cell biological entities, has traditionally been performed in fluorescence. However, it turns out that viruses and virus-like particles such as X31 virus, H1N1, Zika, ebola and SV40, which could range from dimensions of about 20\,nm to beyond 200\,nm \cite{MiloRonPhillips2015}, can be easily detected via iSCAT. This has been demonstrated in environments such as microfluidic channels \cite{Ignatovich2006a,Mitra2010b}, on synthetic lipid bilayers (see Fig.\,\ref{fig:virus}a) \cite{Ewers2007, Kukura2009a}, and on dielectric substrates \cite{Daaboul2010,Daaboul2017}. Simultaneous iSCAT and fluorescence tracking of a virus and a quantum dot on its surface made it possible to visualize not only the nanoscopic binding domains on the membrane but also the rocking and tumbling motion of single viruses  (see Fig.\,\ref{fig:virus}a)\cite{Kukura2009a}. Moreover, iSCAT has been used to uncover to nanometer precision the position and orientational configuration of bacteriophages interacting with a surface as well as to resolve, to a precision of 4200 basepairs, the kinetics of DNA ejection following stimulation \cite{Goldfain2016} - shown in Fig.\,\ref{fig:virus}b. These activities have also motivated attempts to detect pure scattering of single viruses in conventional dark-field \cite{Weigel2014b} and through light scattering from a nanofluidic channel \cite{Faez2015b}.

The large iSCAT signal of most viruses opens the way to label-free high-speed, high-resolution and long-term imaging of single viruses and quantitative study of their interactions with cells and cellular environment \cite{Huang2017a}. iSCAT imaging of viral interactions is still in its infancy and is somewhat slowed down by the difficulty of integrating biosafe conditions into conventional optical laboratories, but it promises to provide invaluable insight about the secrets of viruses, which contain strong inherent inhomogeneities in their structure and function. 

\subsubsection{Proteins}
Proteins are omnipresent in our body, taking on critical roles in essentially every step of our physiology. They are responsible for the function, structure and regulation of our organs and tissue, as well as performing many of the functions of the cell. For these reasons, study of single proteins presents an enormously instructive, but challenging, prospect. 

Proteins range in molecular weight from a few to a several hundred kDa. The ubiquitous protein albumin, for example, has a molecular weight of 65\,kDa. The size of proteins lies in range of a few nanometers, and considering that the effective refractive index of biological matter does not vary much about a value of 1.5, the iSCAT signal of a typical protein can be more than 1000 times smaller than that of a typical virus. Nevertheless, the scattering cross-section of a protein such as albumin turns out to be in the order $2.5\times10^{-15}$\,cm$^{-2}$ at 280\,nm. 

While detection of single dye molecules via iSCAT was performed in a focused laser beam, we had to adopt a wide-field imaging scheme for developing a practical biosensing platform. Here, we focused the incident laser beam in the back focal plane of the microscope objective to achieve uniform illumination of an area with lateral extension in the order of 5-10\,$\mu$m and used CMOS cameras for imaging. Aside from simultaneous study of many particles, an advantage of a camera as detector is that the total power recorded over all pixels can be conveniently used to register the incident power and thus account for its fluctuations. Moreover, recent camera technologies easily allow imaging at speeds of several tens of kHz up to MHz, which is far faster than scanning schemes, whether they use piezoelectric actuators \cite{Lindfors2004, Ewers2007,Celebrano2011a} or acousto-optical deflectors \cite{Kukura2009a,OrtegaArroyo2014f}.

\begin{figure}[!phtb]
    \centering
   \includegraphics[scale=.6]{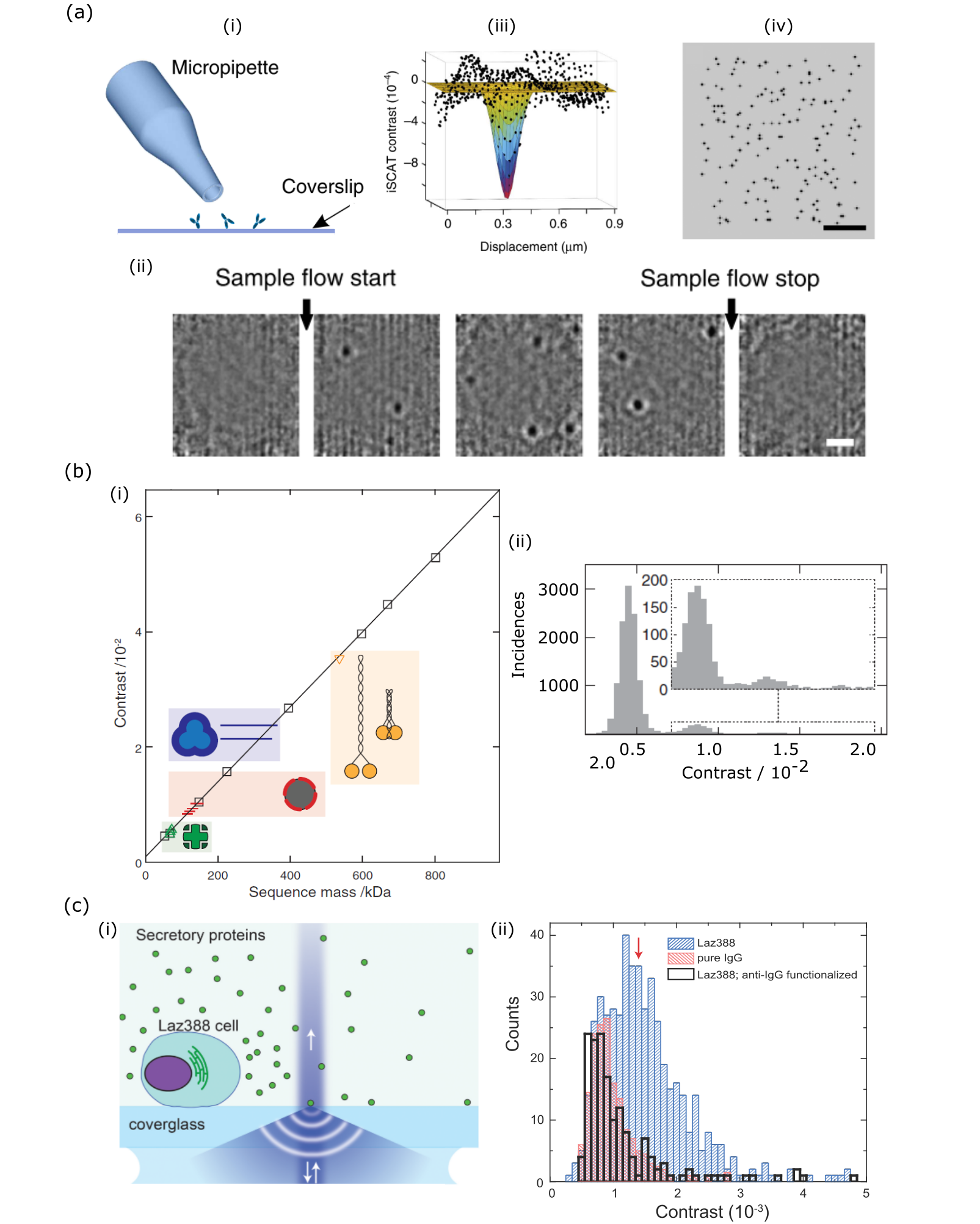}
    \caption{Detection of single protein extinction. (a) Schematic for the first demonstration of single protein detection: (i) A solution of proteins is pipetted upon a functionalized cover slip. (ii) Background-corrected differential iSCAT images showing the cover-slip as pipette flow begins and ends, depositing single proteins which appear as dark spots. (iii) A super-localizing fit to one protein PSF to resolve individual bindings sites, which can be repeated for all bindings (iv) \cite{Piliarik2014}. Reproduced with permission from Nature Publishing Group. Mass calibration (b): The contrast of single and complexed proteins can be calibrated to their mass (i), with monomeric and oligomeric derivatives also resolvable by this technique (ii) \cite{Young2018}. Reproduced with permission from the American Association for the Advancement of Science. (c) Single-cell secretion dynamics resolved at single-protein sensitivity with iSCAT. (i) Schematic of the experimental arrangement wherein a secreting cell is positioned near to the iSCAT field of view, the contrast distribution of proteins secreted by the cell as they bind to an anti-IgG functionalized surface (blank) agrees with that expected from pure IgG (red). iSCAT also reveals abundance of other secreted products (blue) \cite{McDonald2018}. Reproduced with permission from the American Chemical Society.} 
    \label{fig:single-protein}
\end{figure}

The first sensing experiments were reported in 2014 on single proteins as light as about 50\,kDa \cite{Piliarik2014}, shown in Fig.\,\ref{fig:single-protein}a. A careful analysis of the obtained iSCAT contrast for several proteins of different size and mass as well as comparison with single-molecule fluorescence measurements demonstrated the potential of iSCAT for label-free detection of single proteins. This method has several decisive advantages over other biosensing solutions. First, the ability to count single proteins brings sensing to its absolute limit, secondly this happens over a large surface area as opposed to methods relying on plasmonic antennas \cite{Zijlstra2012a} or optical microcavities \cite{Vollmer2008} with very limited active area, thirdly imaging provides invaluable information about the spatial distribution and position of each protein, and finally the essential setup is very simple. As in the great majority of biosensing platforms based on surface plasmons, mechanical oscillators, or microcavities, however, specificity has to be reached via surface functionalization. 

The linearity of the iSCAT contrast with protein mass allows classification of the detected proteins according to their size (see Fig.\,\ref{fig:single-protein}b). This feature has been recently used to demonstrate the application of iSCAT to quantitative mass spectrometry \cite{Young2018}. Moreover, this work nicely shows the ability of iSCAT to watch dynamics of molecular processes, such as protein aggregation, cross-linking and oligomerization. Indeed, iSCAT has been successfully applied to a range of related investigations such as self-assembly of individual tubulin dimers to a growing microtubule \cite{Mickolajczyk2018}, disassembly of a single microtubule \cite{Andrecka2016}, real-time monitoring of 28\,nm viral capsid self-assembling around a viral RNA scaffold \cite{Garmann2018}, and the growth, attachment and retraction of bacterial pili \cite{Tala2018}.  

Another recent application of label-free single-protein detection was showcased in the context of real-time investigation of cellular secretion \cite{McDonald2018,Gemeinhardt2018}, shown in Fig.\,\ref{fig:single-protein}c. Secretion is the basis of intercellular communication and has been a subject of single-cell studies using different methods \cite{Heath2016}, however single protein sensitivity was only possible with iSCAT. In a proof of principle experiment, immunoglobulin G (IgG) antibodies of mass 150\,kDa (4\,nm size) were detected in a spatio-temporally resolved fashion following secretion by Laz388 cells \cite{McDonald2018,Gemeinhardt2018}. This study paves the way for future experiments in a wide range of studies, e.g. the interaction of immune cells. iSCAT secretion has also been recently extended to plasmonic substrates, where the incident light is coupled to surface plasmons in a thin gold surface \cite{Yang2018}.

\subsection{Dynamics in nanobiology}
High speed, high spatial precision and a long measurement duration in imaging are some of the key advantages of iSCAT that become particularly important when investigating processes such as diffusion and transport of viruses, proteins, lipids or other nanoscopic entities. In this section, we present a few case studies where iSCAT was used in this context. 

\subsubsection{Protein tracking} 
\begin{figure}[!htb]
    \centering
   \includegraphics[scale=.65]{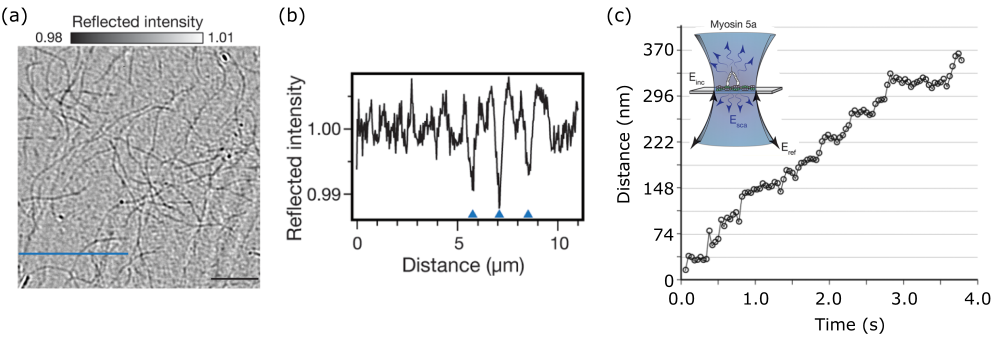}
    \caption{Single protein tracking on an actin filament. (a) iSCAT image of unlabeled actin filaments adhered to a cover-slip. Scale bar denotes $5\,\mu$m. (b) Signal intensity for the blue line marked in (a). Blue arrow heads denote three actin filaments. (c) Motor protein myosin\,5a walking velocity along a single actin filament when tracked by iSCAT. Frame rate is 25\,Hz after temporal averaging. Inset: schematic of specimen detection \cite{OrtegaArroyo2014f}. Reproduced with permission from the American Chemical Society. }
    \label{fig:actin}
\end{figure}

In the previous section, we discussed the power of iSCAT in detecting unlabeled proteins. The precision and fast temporal imaging of iSCAT microscopy also lends itself to investigation of the \textit{mobility} of single proteins such as mysosin-5 motor protein on actin \cite{OrtegaArroyo2014f, Andrecka2015} (see Figure \ref{fig:actin}). In addition, the mobility of single microtubules can be investigated to high precision when forming a gliding assay upon kinesin \cite{Andrecka2016}. Furthermore, motion of unlabeled small proteins upon landing on a surface has been visualized using iSCAT \cite{Spindler2016}.

\subsubsection{Lipid membranes} 
Lipid membranes are one of the most important substrates in biology, often composed of a bilayer of phospholipids molecules and integral and peripheral proteins of dynamic composition. The membrane functions as a selectively permeable barrier and a catalytic reaction site which gates cellular function and communication. Although the ultimate goal is the understanding of full cellular membranes, investigation into the mobility and statistical physics of diffusion within well-defined biophysical model systems is highly instructive and is an active area of research. 

Previous efforts to investigate diffusion in membranes, stretching back to the 1970s and still active today, have depended upon fluorescence labeling, whether that be for ensemble measurements such as in fluorescence recovery after photo-bleaching (FRAP), fluorescence correlation spectroscopy (FCS) and its resolution-enhanced version STED-FCS or alternatively single-molecule tracking. The spatial and temporal resolution to which one can investigate diffusion using these strategies are throttled by the limitations within the photophysics of fluorescence, namely a finite and low yield of emitted photons as well as blinking and bleaching behavior. 

\begin{figure}[!phtb]
    \centering
   \includegraphics[scale=.7]{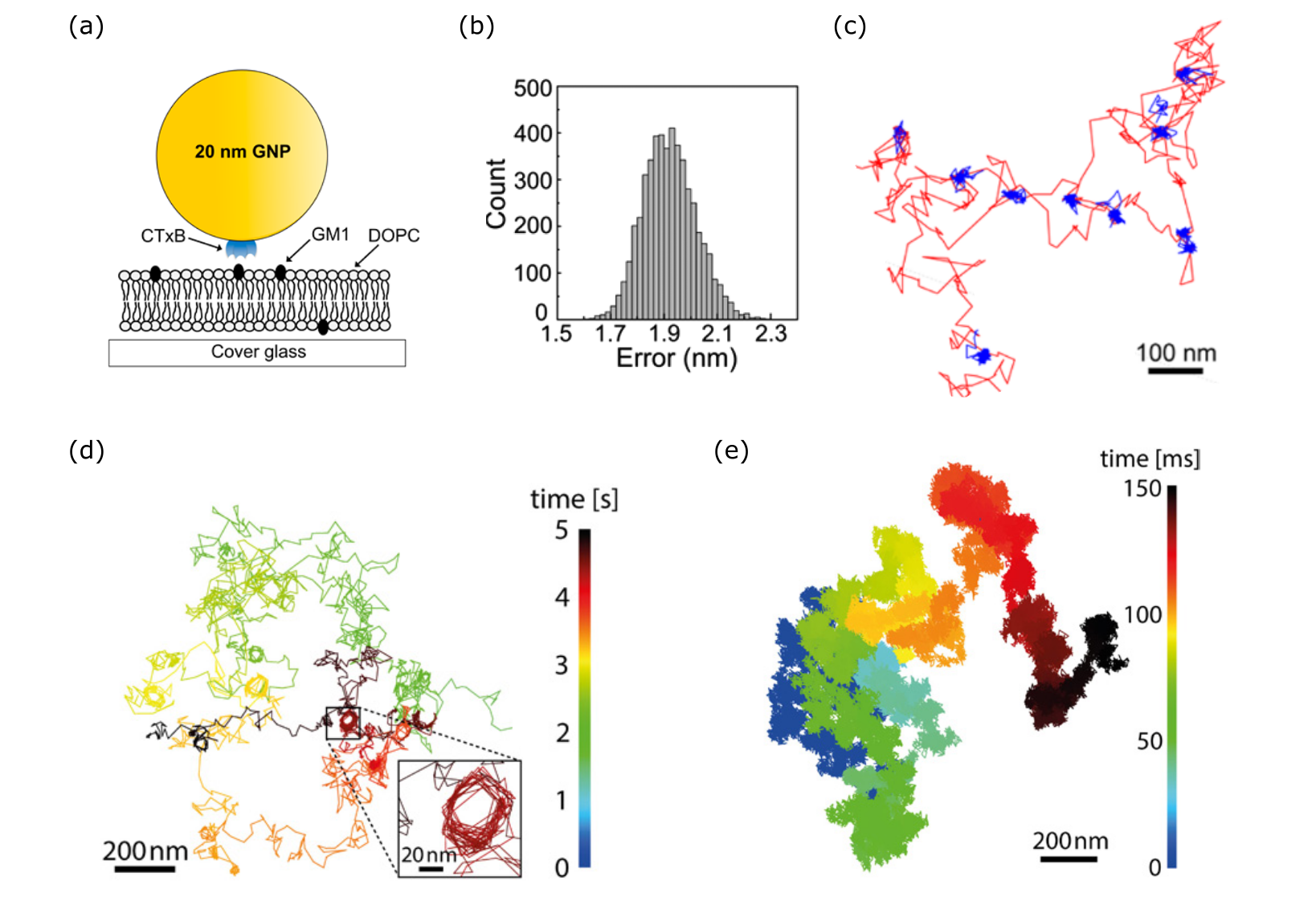}
    \caption{High spatio-temporal resolution of lipid diffusion in a bilyer membrane. (a) Schematic of a GNP tagging a GM1 lipid within a synthetic bilayer membrane. (b) Lateral localization of the GNP probe to nanometric precision. The diffusive trajectory of the mobile lipid (c) reveals regions of transient confinement (marked blue), features resolved by the nanometric precision and 1\,kHz temporal resolution \cite{Hsieh2014}. Reproduced with permission from the American Chemical Society. (d) Additionally, trajectories can reveal circular nanoconfinements (inset), trajectory duration 5\,s. (e) Ultra-fast temporal resolution: the diffusive trajectory of a 20\,nm GNP bound to a biotinylated DOPE lipid in a DOPC bilayer, recorded at 913,000\,fps with duration 0.152\,s \cite{Spindler2016}. Reproduced with permission from IOP Publishing.}
    \label{fig:lipid-plane}
\end{figure}

After its early studies of viruses bound to supported lipid bilayers, iSCAT was extended to tracking the diffusion of lipids labeled by GNPs. For example, a CTxB-tagged 20\,nm GNP probe was tethered to DOPE lipids or GM1 gangliosides mixed at low concentration in DOPC formed on glass substrates \cite{Hsieh2014}, illustrated in Fig.\,\ref{fig:lipid-plane}. The high resolution achieved (2\,nm precision, 1000\,fps) led to the observation of mixed mobility of lipid diffusion as well as identification of transient nanoscale confinements \cite{Hsieh2014,Spindler2016}. The curious confinement modes were attributed to the molecular pinning and inter-leaflet coupling between lipid tail domains as the source of the observed transient immobilization on the millisecond time scale \cite{Spillane2014b}. Quantitative iSCAT studies have also enabled clear discrimination of the varying mobility of a lipid diffusing between differently ordered phases within a membrane \cite{Wu2016}. Furthermore, the extended duration over which measurement can be performed has inspired researchers to develop new statistical models to interpret the new generation of experimental results \cite{Slator2018}. It has to be born in mind, however, that quantitative comparisons of diffusion performed by different methods remains a challenge since each labeling approach might introduce a certain systematic bias, and indeed, diffusion coefficients obtained from different techniques might vary \cite{Reina2018}. 

We point out that aside from the nanometric precision to which the particle can be localized in space, iSCAT microscopy enables ultra-short exposure times, granting temporal resolutions that have thus far approached one million frames per second  \cite{Lin2014,Spindler2016}, which for a 20\,nm GNP gives a 10\,nm localization precision. The sheer density of positional information of iSCAT trajectories permits robust statistical analysis of transient behavior contained within. Here, it is important to realize that a high localization precision resulting from long integration times is meaningless in particle tracking if the dynamics at hand are faster than the imaging speed so that the positional information becomes smeared. 

As well as measuring the diffusive properties of membrane constituents, the sensitivity to which iSCAT can image small and faint microscopic entities can be used in imaging the formation of lipid membranes in a similar \textit{label-free} fashion. Here, it is helpful to note that small anunilamellar vesicle (SUV) has already a sufficiently large polarizability to be comfortably detected in iSCAT \cite{Krishnan2010}. This high sensitivity can be exploited to observe, in real time, the docking and rupture of SUVs with a size down to 20\,nm \cite{Andrecka2013,Spindler2016} and co-existing dynamic phases within a supported membrane \cite{DeWit2015}.

\begin{figure}[!htb]
    \centering
   \includegraphics[scale=.65]{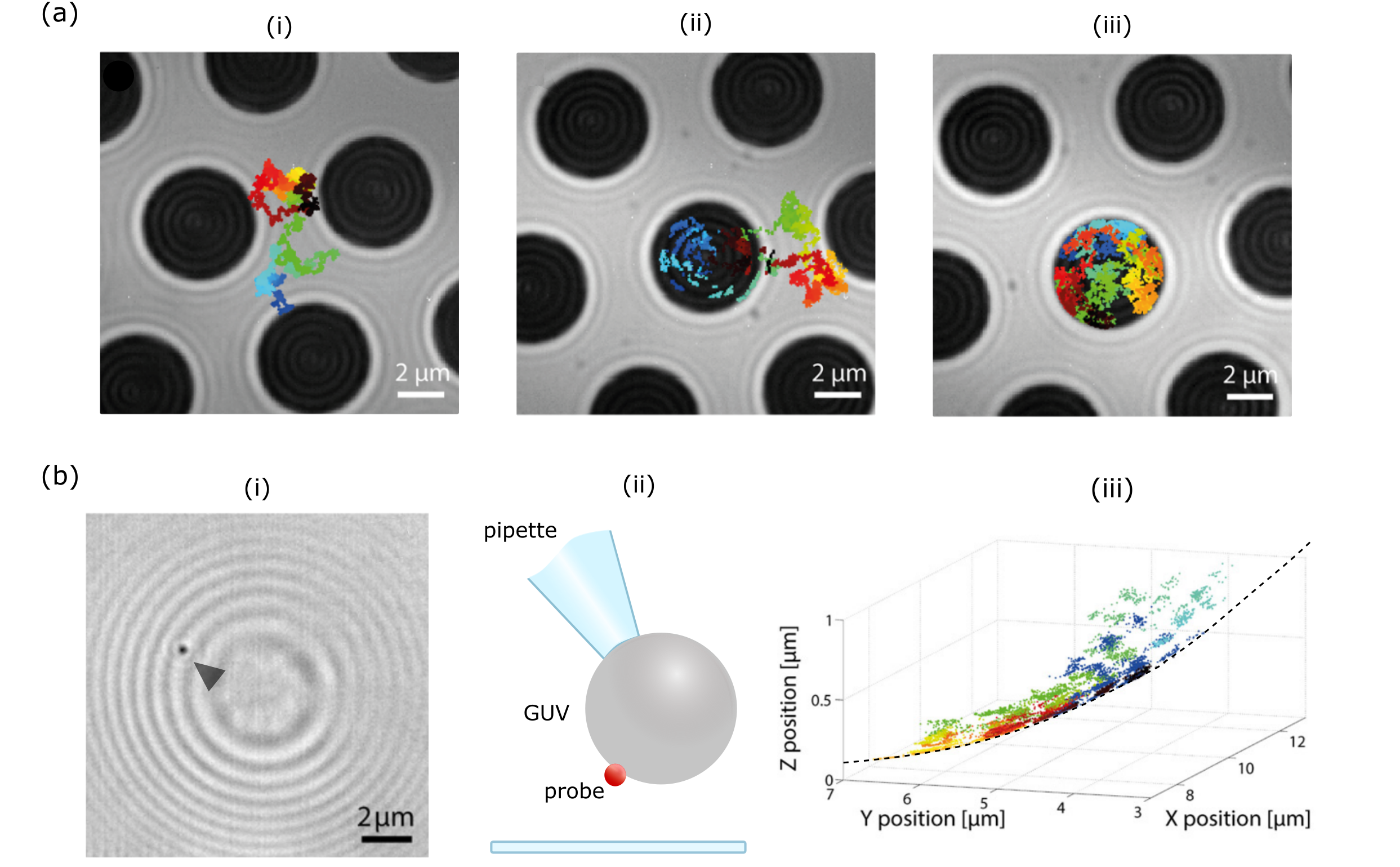}
    \caption{iSCAT microscopy of non-planar model lipid systems. (a) Pore-spanning membrane wherein the synthetic bilayer covers a 5\,$\mu$m hole-filled substrate, such that the membrane is supported (bright contrast) or free-standing (dark contrast). Show are trajectories for a GNP-tagged biotinylated-DOPE lipid showing different mobilities depending on whether the membrane is supported (i) or free-standing (iii), with a change in mobility evident upon region transition \cite{Spindler2018}. Reproduced with permission from the American Chemical Society. (b) Giant unilamellar vesicles (GUVs) provide a completely uniform, substrate-free membrane. (i) Wide-field reflection iSCAT image of the GUV, revealing a Newton ring-like pattern. (ii) A schematic of a GUV held by a pipette during iSCAT imaging, such as that shown in (i). A 10\,s-long trajectory of the Tat virus-like particle which faithfully reproducing the great spherical topology of the GUV \cite{Spindler2016}. Reproduced with permission from IOP Publishing.}
    \label{fig:lipid-GUV}
\end{figure}

Whilst the planar bilayer membrane serves as a convenient model system, concerns over the perturbative influence of the substrate motivate efforts to investigate free-standing model membranes. One such system is the pore-spanning membrane \cite{Spindler2018}, where a continuous lipid membrane spans an array of micron-sized pores, providing regions of supported and free-standing membranes, shown in Fig.\,\ref{fig:lipid-GUV}. High-resolution iSCAT trajectories from GNP-tagged lipids revealed that nanoscale transient confinements were only observed on the supported regions, seemingly confirming suspicions as to the influence of the substrate raised in earlier work \cite{Spindler2018}. 

Another option for substrate-free membrane studies is to work with giant unilamellar vesicles (GUVs) with diameters in the range of tens of micrometers \cite{Stein2017}, which have been pursued as a platform for a minimal cell model \cite{luigi-book}. Preliminary results have been reported on tracking of the three-dimensional diffusion of viral-mimetic particles on a GUV surface, where the particle could be localized to nanometer precision in all dimensions over an extended range \cite{Spindler2016}, illustrated in Fig.\,\ref{fig:lipid-GUV}b-d.

\subsubsection{Imaging cells and associated elements} 

\begin{figure}[!htb]
    \centering
   \includegraphics[scale=.67]{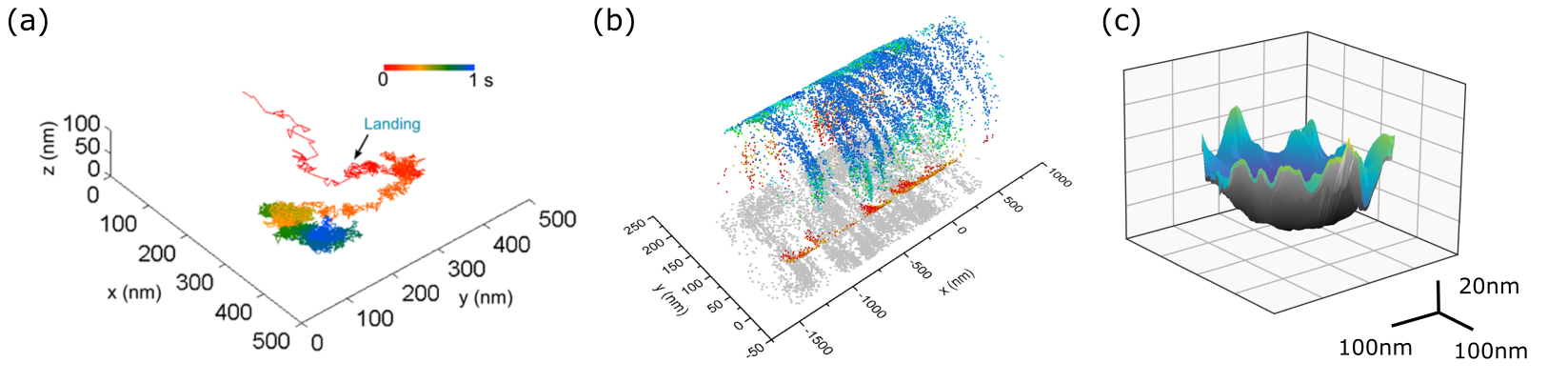}
    \caption{Imaging of cellular features with iSCAT. (a) 3D high-speed tracking of a vaccinia virus landing upon the surface of a live HeLa cell \cite{Huang2017a}. Reproduced with permission from American Chemical Society. ba) 3D diffusion of a GNP on a live neurite \cite{DeWit2018}. Reproduced with permission from Elsevier. (c) High-speed tracking of protein-labeled GNP diffusion within a pit on a live HeLa cell, which when interpolated renders a bowl-like surface \cite{Taylor2018}.}
    \label{fig:cell-imaging}
\end{figure}

Three-dimensional monitoring of molecules on the plasma membrane of live cells to high speed and precision stands as one of the most exciting challenges to unravel in cell biology \cite{BernardinodelaSerna2016,Sezgin2017a}. To date, nanoscale transient organization has precluded satisfactory or compelling investigation by fluorescence methodologies, which fall short in many respects, especially in capturing three-dimensional landscapes. While iSCAT is predestined for addressing this issue, its ultrahigh sensitivity virtue is accompanied by the vice of a large speckle-like pattern originating from the cellular membrane and corpus, thus making cellular iSCAT particle tracking a challenge. Nevertheless, first attempts in tackling this issue have been successfully reported \cite{Huang2017a,DeWit2018,Taylor2018} (see Fig.\,\ref{fig:cell-imaging}). In particular, very fast detection allows one to reach nanometric-microsecond three-dimensional tracking of a transmembrane protein on the live HeLa cell, unraveling details such as heterogenous mobility of the protein, confinement into clathrin-like lattices and extended-duration directed diffusion along filopodia \cite{Taylor2018}.

When imaging super-wavelength objects such as cells, cell nuclei or bacteria, one no longer speaks of scattering, but rather of reflection, absorption and transmission, as is common in elementary textbooks. As pointed out in the introductory section, interferometric microscopy has a long and rich history. In the past decade, however, iSCAT and its related techniques have ushered in a revival of interferometric imaging for cell biology applications, such as imaging microtubules \cite{Simmert2017,Mahamdeh2018,Wang2011,Kandel2017}, actin \cite{OrtegaArroyo2014f}, the acto-myosin network \cite{Koester2017a}, as well as organelles and micron or smaller-sized vesicle containers which assist in the transport of material throughout the cell \cite{Huang2017b}. Echoing previous efforts from the holographic community \cite{Choi2007,Yaqoob2011a}, interferometric efforts are now also focusing on identifying the small changes in the cell membrane, where previously the whole cell was profiled. Examples include the progression of ideas from holography to gradient light interference microscopy \cite{Nguyen2017}, recent efforts to profile membrane adhesion sites and topology in both wide-field \cite{Klein2013} and confocal-scanning reflection interference microscopy \cite{Matsuzaki2014b} or wide-field iSCAT \cite{Park2018}.

Similarly, there have been increasing efforts to harness interferometric imaging to investigate the mechanical properties of the cell membrane. Building upon initial efforts from quantitative phase imaging \cite{Popescu2006a,Popescu2005b}, recent works have sought to turn attention to fast and nanometer-level dynamics that require high sensitivity to elucidate inherent fluctuations \cite{Monzel2016, Biswas2017, Yu2018,Yang2018a}. Results provide insight into the mechanical properties or changes in the membrane following the execution of action-potentials and sub-nanometre-level twitching across the neuronal cell in response to stimulation.

Finally, it is interesting to note that some of the early efforts of the 1990s actually used iSCAT in transmission mode to track membrane proteins in living cells \cite{Kusumi1993b} although the underlying physics was not formulated, and thus, the virtues of iSCAT were not harnessed to reach the quality of trajectories that are accessible today. 

\subsubsection{Other emerging applications} 

More recently, iSCAT imaging was also used to visualize excitation and relaxation of charge-carriers within various semiconductor and organic crystals, following ultrafast pumping \cite{Delor2018}. The changes in local carrier density (refractive index) induced by the pump excitation mark the flow of energy in the material which are visualized as local changes in contrast, with contrast sensitivity down to $10^{-4}$. The relaxation dynamics could thus be imaged through microscopy, and the role of grain boundaries and material anisotropy were explored. In the wake of such efforts, we also anticipate great utility of this technique in the field of atomic physics.

\section{Summary \& Outlook} 
We began this review with a historical overview of different microscopy modalities where interference plays a central role. In particular, we emphasized that the physical mechanism behind conventional bright-field microscopy is extinction, which follows from formulating an interference problem between the illumination and the field that has interacted with the object. Methods such as phase contrast microscopy, differential interference microscopy, various versions of interference reflection microscopy and digital holography all share the same fundamental physics and only differ in the way phases, path difference and index modulations are introduced in the problem. The main focus of the review has been to discuss the latest developments of interferometric scattering (iSCAT) microscopy upon the past decade and a half for detecting, sensing, imaging and tracking \textit{nanoparticles} \cite{Lindfors2004,Jacobsen2006a,Ewers2007,Kukura2009,Kukura2009a}. Although the equation describing iSCAT uses very much the same interference expression known in conventional interferometric microscopy, its application to nanoparticles was indeed new and has led to a wealth of recent studies. These have often been presented under new acronyms inspired by variations of the illumination and detection schemes. In this work, we have aimed at unifying these efforts under a common theoretical and experimental roof and have attempted to present a concise discussion of some case studies and various implementations of iSCAT that are rapidly emerging. 

While in its infancy, we believe that iSCAT microscopy stands as an exciting next-generation tool for uncovering the nanobiology of the cell membrane as well as advanced diagnostics and laboratory analytics applications. The sensitivity and performance of iSCAT is only limited by the signal-to-noise ratio, which in turn depends on factors such as the choice of detector and background fluctuations that result from specimen dynamics. Advances in detector technologies, image processing and machine learning, as well as optical techniques for imaging through scattering media, promise to pave the path to these goals. Thus, we are convinced that the limits of iSCAT will be continually pushed in the near future, exploring unlabeled proteins well under 20\,kDa not only in a biosensor geometry but also in the living cell membrane. The simplicity of an iSCAT microscope lends it to miniaturization into a compact instrument as well as integration into existing microscope stations such as a laser scanning confocal microscope, thus providing prospects of very wide-spread use.

\textbf{Acknowledgements:} We are grateful to a large number of fantastic group members, former and present, who have contributed to the advances of iSCAT in our laboratories. We also thank the Alexander von Humboldt Foundation for their generous support in the context of a Humboldt Professorship (VS) and Postdoctoral Fellowship (RWT) as well the Max Planck Society for continuous support.

\end{document}